\documentclass[12pt]{article}
\usepackage{epsf}
\begin{document}
\newcounter{numrow}
\newcommand{\poczn}{
\stepcounter{equation}
\setcounter{numrow}{\value{equation}}
\setcounter{equation}{0}
\renewcommand{\theequation}{\thesection.\thenumrow\alph{equation}}}
\newcommand{\konn}{
\setcounter{equation}{\value{numrow}}
\renewcommand{\theequation}{\thesection.\arabic{equation}}}
\renewcommand{\theequation}{\thesection.\arabic{equation}}
\newcommand{\IH}{\relax{\rm I\kern-.18em H}}
\thispagestyle{empty}
\title{Irreversible Quantum Mechanics in the Neutral K-System}
\author{Arno Bohm\thanks{\hspace{8pt}e-mail:  bohm@physics.utexas.edu}\\
\sl{University of Texas at Austin}\\
\sl{Austin, Texas   78712}}
\date{}
\maketitle
\begin{abstract}
The neutral Kaon system is used to test the quantum theory of
resonance scattering and decay
phenomena.  The two dimensional Lee-Oehme-Yang theory with complex
Hamiltonian is
obtained by truncating the complex basis vector expansion of the
exact theory in Rigged
Hilbert space.  This can be done for $K_1$ and $K_2$ as well
as for $K_S$ and $K_L$, depending
upon whether one chooses the (self-adjoint, semi-bounded)
Hamiltonian as commuting or
non-commuting with CP.  As an unexpected curiosity one can
show that the exact
theory (without truncation) predicts long-time $2\pi$ decays
of the neutral Kaon system even
if the Hamiltonian conserves CP.
\end{abstract}
\newpage
\section{The standard two-dimensional
effective theory with complex
Hamiltonian}

The phenomenological characteristics of a resonance or a decaying
particle are
its energy $E_R$ (resonance energy) or
relativistically its mass $m$, and its
width $\Gamma=\hbar/\tau_R$ where $\tau_R$ is its lifetime
(in its rest frame).
These two real numbers are combined into a complex energy
$z_R=E_R-i{\gamma\over2}$ $\bigl($(or relativistically into
a complex value
$s_R=(m-i{\gamma\over2})^2$ of the Mandelstam variable $s$,
which is the
(energy)$^2$ in the rest frame of the decaying state or the
(total energy)$^2$ in
the center of mass frame of the decay products (often not
$m$ but $M$ defined by
$s_R=M^2-iM\gamma$ is called the mass of the resonance)$\bigr)$.
This complex energy
$z_R$ or complex mass-squared $s_R$ can be  defined as the position
of the first
order (resonance) pole in the second sheet of the analytically
continued
$S$-matrix, in the same way as stationary states are given by
bound state poles of
the $S$-matrix.  Empirically, stability or the value of the
lifetime does not appear
to be a criterion for elementarity.  Stable particles
are not qualitatively different
from quasistable particles, but only quantitatively
different by a zero or very
small value of $\Gamma$.   (A particle decays if it can
decay and it is stable
if selection rules for some quantum numbers prevent
it from decaying.)
Therefore both stable and quasistable states should be
described on the same
footing, as is the case if one defines them by $S$-matrix poles.

Since a stationary state characterized by
a real energy value $E_s$ is
described not only by a bound state pole on the negative real axis but also
by an eigenvector of the (self-adjoint, semi-bounded)
Hamiltonian $H$ with
eigenvalue $E_s$, a ``pure" decaying state should
also be described by an
``eigenvector" of $H$ but with complex eigenvalue $z_R$.
In the standard,
Hilbert space formulation of quantum mechanics such vectors do not exist.
But since it is of  practical importance for the  phenomenological
description of
experiments to have a vector space  description, this
deficiency of the Hilbert space quantum mechanics
has not prevented the
practitioners  from using eigenvectors with complex energy in
``phenomenological",
``effective" theories of decay.$^{1)}$
The Lee-Oehme-Yang theory for the time evolution of the
two-resonance neutral Kaon system is the most celebrated
example$^{2)}$ of this.

The Lee-Ohme-Yang theory is usually justified as the Weisskopf-Wigner
approximation applied to the two Kaon states $|K^0\rangle,|\bar
K^0\rangle$, in which the neutral $K^0$-system is prepared in an
inelastic scattering experiment by strong interaction (e.g.
$\pi^-p\to\Lambda K^0$).

One then introduces the two linear combinations $|K_1\rangle$ and
$|K_2\rangle$ (leaving aside the CP violating Hamiltonian) ---
or the $|K_S\rangle$ and
$|K_L\rangle$
(if one includes CP violation in the Hamiltonian):
\poczn
\begin{eqnarray}
&|K_S\rangle={N_S\over\sqrt{2}}
\left((1+\epsilon_S)|K^0\rangle+(1-\epsilon_S)|\bar
K^0\rangle\right)\approx|K_1\rangle\phantom{0}
{\rm for}\ \epsilon_S\to0\\
&|K_L\rangle={N_L\over\sqrt{2}}
\bigl((1+\epsilon_L)|K^0\rangle-(1-\epsilon_L)|\bar
K^0\rangle\bigr)\approx|K_2\rangle\phantom{0}
{\rm for}\ \epsilon_L\to0
\end{eqnarray}
\konn
The $|K_1\rangle$ and $|K_2\rangle$
(or $|K_S\rangle$ and $|K_L\rangle$)
are defined
to be eigenstates of the complex
``effective" Hamiltonian $H=M-{i\over2}\Gamma$.
These vectors, and not the $|K^0\rangle,|\bar K^0\rangle$,
are
assumed to represent the particle states,
with the justification that ``since we
should properly reserve the name ``particle"
for an object with unique
lifetime,
$K_1$, and $K_2$ are the true particles".$^{3)}$
This effective Hamiltonian we
call $H$ if the CP  violating term $H_{sw}$ is left aside,
and we call it
\begin{equation}
\IH=H+H_{sw}=H_0+H_w+H_{sw}=H_0+\IH_{int}\,\,\,{\rm with}\,\,\,
[H_{sw},CP]\not=0
\end{equation}
if {\bf s}ome {\bf w}eak additional
CP-violating interaction Hamiltonian $H_{sw}$ (not
necessarily superweak) is included.
$H$ is assumed to be normal
(i.e. $M,\Gamma$
are hermitian and commute, which excludes Jordan blocks),
and thus it is diagonalizable with
eigenvalues given by
\poczn
\begin{equation}
H|K_1^{0\thinspace-}\rangle=
(m_1-i{\gamma_1\over2})|K^{0-}_1\rangle\quad;\quad
H|K^{0-}_2\rangle=(m_2-i{\gamma_2\over2})K^{0-}_2\rangle\ ,
\end{equation}
and similarly for ${\IH}$
\begin{equation}
{\IH}|K_S^{\enspace-}\rangle=(m_S-i{\gamma_S\over2})|K_S^{\enspace-}
\rangle\quad;\quad
{\IH}|K_L^-\rangle=
(m_L-i{\gamma_L\over2})|K_L^{\enspace-}\rangle\ .
\end{equation}
The time evolution for $t\geq0$ is given by
\stepcounter{numrow}
\setcounter{equation}{0}
\begin{eqnarray}
e^{-iHt}|K^{0\enspace-}_1\rangle=e^{-im_1t}
e^{-{\gamma_1\over2}t}|K_1^{0\enspace-}\rangle\\
e^{-iHt}|K^{0\enspace-}_2\rangle=e^{-im_2}
e^{-{\gamma_2\over2}t}|K_2^{0\enspace-}\rangle\nonumber
\end{eqnarray}
or, if \IH\  is the Hamiltonian, by
\begin{eqnarray}
e^{-i{\IH}t}|K_S^{\enspace-}\rangle=
e^{-i(m_S-{i\gamma_S\over2})t}|K_S^{\enspace-}\rangle\\
e^{-i{\IH}t}|K_L^{\enspace-}\rangle=
e^{-i(m_L-{i\gamma_L\over2})t}|K_L^{\enspace-}\rangle\nonumber
\end{eqnarray}
\konn
The entire effective theory of the neutral
$K$-meson-system then takes place in
this 2-dimensional space ${\cal H}_2$ spanned
by the eigenvectors (3).  An
arbitrary coherent mixture (superposition) in the
$K^0$-beam is given by
\begin{equation}
\phi^{eff}(0)=|K_S^-\rangle a_S+|K_L^{\enspace-}\rangle
a_L\propto|K_L^-\rangle+\rho|K_S^-\rangle
\end{equation}
It has the time evolution:
\begin{equation}
\phi^{eff}(t)=e^{-i\IH
t}\phi^{eff}=e^{-im_St}
e^{-{i\gamma_S\over2}t}|K_S^{\enspace-}\rangle
a_S+e^{-im_Lt}e^{-{i\gamma_2\over2}t}
|K_L^{\enspace-}\rangle a_L
\end{equation}
(or a corresponding expansion in terms of $K_1,K_2$, with
$a_1, a_2$, if (3a) and (4a)
hold).  The $|K_S\rangle$ and $|K_L\rangle$
in the coherent (pure) beam
state (5) are  conventionally
expressed in terms of $|K^0\rangle$ and
$|\bar K^0\rangle$ by (1),
though the $|K^0\rangle,|\bar K^0\rangle$ and the
$|K_S^{\enspace-}\rangle,\ |K_L^{\enspace-}\rangle$
belong to different Hamiltonians
$H_0$ and
$\IH$ (or
$H_0$ and
$H$ for $H_{sw}=0$), respectively.
The $|K^0\rangle$  and $|\bar K^0\rangle$ should, therefore,
span a space different from
${\cal H}_2$.   We annotate this difference between these
two kinds of vectors by the superscript
$^-$ in
$|K_2^{\enspace-}\rangle$ etc.
We shall discuss this notation in more detail in section 3
below.   An incoherent mixture,
which one usually encounters experimentally in the initial
$K^0$ beam$^{4)}$ (or which arises from a pure state if one
uses a theory based on a Liouville
equation for the neutral $K$-system$^{5)}$),
is described by a density matrix
or a statistical operator in the space ${\cal H}_2$.

The quantity that one always considers$^{6)}$ is the
instantaneous decay rate of the $K^0$
state at time $t$ into the detected decay channel $c$.
This transition rate is proportional to
$2\pi\rho_c|<c|\IH_{int}|\phi(t)\rangle|^2$, where $c=\pi^+\pi^-$ or
$\pi^0\pi^0$ or any other decay channel.
$\rho_c$ is the phase space factor (density of states in
channel
$c$) and
$|c\rangle$ is the eigenvector of the interaction-free
Hamiltonian $H_0$.
The amplitude of this rate is according to (6) given by
\begin{eqnarray}
\langle\pi\pi|\IH_{wt}|\phi^{eff}(t)\rangle&=&
\langle\pi\pi|\IH_{int}|K_S^-\rangle a_Se^{-im_St}
e^{-{\gamma_L\over2}t}\nonumber\\
&&+\langle\pi\pi|\IH_{in}|K_L^-\rangle a_Le^{-im_St}
e^{-{\gamma_L\over2}t}
\end{eqnarray}
One usually considers the ratio
\begin{equation}
R(t)\equiv{|\langle\pi\pi|\IH_{int}|\phi^{eff}(t)
\rangle|^2
\over|\langle\pi\pi|\IH_{int}|K_S\rangle|^2}
\end{equation}
for which one obtains from (7)
\begin{equation}
R(t)=|a_S|^2e^{-\gamma_St}+|a_L\eta|^2e^{-\gamma_Lt}
+2|a_S||a_L||\eta|e^{-(\gamma_L+\gamma_S){t\over2}}
\cos(\Delta mt+\varphi)
\end{equation}
where
\[
\Delta m=m_L-m_S
\]
and where
\begin{eqnarray}
\eta&=
{\langle\pi\pi|\IH_{int}|K_L\rangle\over\langle\pi\pi|\IH_{int}
|K_S\rangle}=|\eta|e^{-i\varphi}\nonumber\\
\varphi&=arg\  a_S-arg\ (\eta a_L)
\end{eqnarray}
If $|\phi^{eff}\rangle$ is
$|K^0\rangle$, then
$a_S=a_L={1\over\sqrt{2}}$.  (If $\phi^{eff}$ is a coherent
mixture behind a regenerator then
$a_S=\rho\ ; a_L=1$.)
For large values of $t\approx20\tau_S=20{1\over\gamma_S}$
($\tau_S=0.893\pm0.001\times10^{-10}\sec$) only the
second term in $R(t=20\tau_S)$ of (9)
does not have a factor of
$e^{-{1\over2}\gamma_St}\approx10^{-5}$ or smaller,
so $R(t)$ is given
by
\begin{equation}
R(t=20\tau_S)={1\over2}|\eta|^2e^{-\gamma_Lt}\enspace,
\quad\gamma_L/\gamma_S\approx1.72\times10^{-3}\ .
\end{equation}
If in (2)
$H_{sw}$ would be zero and if in (1a)
$|K_L\rangle=|K_2\rangle$ with
$CP|K_2\rangle=-|K_2\rangle$, then (due to
$CP|\pi\pi\rangle=+|\pi\pi\rangle)$
$\langle\pi\pi|\IH_{int}
|K_L\rangle\rightarrow\langle\pi\pi|H_w|K_2\rangle=0$, i.e.
$\eta$ in (11) should
be zero.
Experimentally, however,  one observes
$R(t=$ large) $\not=0$ (Princeton effect$^{7)}$).
This is explained by the existence of an $H_{sw}$
with the properties of (2), and by the
decaying particle states
$|K_L\rangle$
and
$|K_S\rangle$
not being the CP eigenstates
$|K_2\rangle$ and $|K_1\rangle$ respectively.

The two complex parameters,
$\eta_{+-}=|\eta_{+-}|e^{i\varphi+-}$, given by (10) for
$\langle\pi\pi|=\langle\pi^+\pi^-|$, and $\eta_{00}$ given by (10) for
$\langle\pi\pi|=\langle\pi^0\pi^0|$,
are the observable quantities in terms of which one
usually expresses the experimental data ascribed to CP violation.
The latest experimental
data, which may$^{8	)}$ or which may not$^{9)}$
indicate direct CP violation
($\langle\pi\pi|\IH_{int}|K_2^{\enspace-}\rangle\not=0,\
\eta_{+-}\not=\eta_{00})$,  give
the following values for the CP-violation parameters:
\begin{equation}
|\eta_{+-}|=(2.269\pm0.023)10^{-3},\enspace
\phi_{+-}=44.3^{\circ}\pm0.8^\circ\ ;\enspace
(\eta_{00}\approx\eta_{+-})\ .
\end{equation}
Inserting this into (11) we obtain the following experimental
value for $R(t)$:
\begin{equation}
R(t=20\tau_s)\approx{1\over2}
|2.27\cdot10^{-3}|^2\cdot 0.966\approx{1\over2}
|2.23\cdot10^{-3}|^2
\end{equation}
This number we shall use for the phenomenological
analysis in section 4.
\setcounter{equation}{0}
\section{
The Rigged Hilbert Space Formulation of Quantum Theory}

Vectors with the properties (1.4) and (1.3) have of course no place in the
Hilbert space of the standard quantum theory,
where all vectors have a unitary
time evolution and where (self adjoint semi-bounded) Hamiltonians cannot have
complex eigenvalues.  If, on the other hand,
one forces the neutral $K$ vectors
into the Hilbert space then one derives all kind of
mathematical consequences for which there exists no experimental evidence,
like deviations
from the exponential decay law, and vacuum regeneration of $K_S$ from
$K_L$.$^{10)}$
The resolution
of these incongruities is, of course, that the two-dimensional space
${\cal H}_2$ of the decaying neutral
$K$-meson system can not be contained in
the Hilbert space.
But we shall show that ${\cal H}_2$ is contained in the dual space
$\Phi^\times$ of a rigged Hilbert space
(or Gelfand triplet) $\Phi\subset{\cal
H}\subset\Phi^\times$.

A theory of resonance scattering and decay
has been developed over the past two
decades$^{11,12)}$ which uses the rigged Hilbert space (RHS)
for\-mu\-la\-tion$^{13)}$
of quantum mechanics.

Whereas in the Hilbert space the solutions of the
Schr\"odinger equation can have only unitary
time evolution and the eigenvalues of self-adjoint
operators can only be real, the RHS
formulation allows for a greater variety of
solutions and for new initial and boundary
conditions with a preferred direction of time (irreversibility).
In this formulation,
heuristic motions like Dirac kets $|E\rangle$, Gamow's
(exponentially decaying ``state")
vectors $|E-i{\Gamma\over2}\rangle$,
and Peierls purely outgoing boundary conditions,$^{1)}$
can be given a unified and mathematically meaningful foundation.
Vectors with the properties
(1.3) and (1.4) are well defined in the RHS formulation as
generalized eigenvectors (cf
Appendix (A6)) of a self-adjoint Hamiltonian
(semi-bounded essentially-self-adjoint operator
$H$ in an infinite dimensional space),
and the time evolution of these vectors is given by an
irreversible semigroup generated by the Hamiltonian.
These new vectors, which can also be obtained
from a resonance pole of the analytically continued
$S$-matrix and consequently have a Breit-Wigner
energy distribution, have been called Gamow vectors.

Since this paper addresses a physics problem, we do not give here the
precise mathematical definition of the RHS in general and of the particular
spaces $\Phi,\ \Phi_+,\ \Phi_-$ and their topological duals (space of
continuous antilinear functionals)
$\Phi^\times,\ \Phi_+^\times,\
\Phi^\times_-$, which we shall use in this paper.
In the appendix A we give a brief and casual
description of the RHS.
For the calculations in this paper we adopt the modus operandi of a
physicist and do not worry about the precise mathematical definitions of
${\cal H}$ or $\Phi$ or
$\Phi_+,\ \Phi_-,\ \Phi^\times_+$ etc.  All these spaces are different
topological completions of the same pre-Hilbert space $\Psi$ ($i.e.$ a linear
space $\Psi$ with ``scalar product" denoted by $(\psi,F)$ or
by
$\langle\psi|F\rangle$).
Here we use the algebraic space $\Psi$ and some additional
rules
which can be justified by the mathematics of the RHS.
These additional rules include
the well known rules of the Dirac bra-and ket formalism,
some mathematical properties of the
Gamow vectors
(e.g. those given by (1.3) and (1.4)) and in particular some basis
vector representations
which were not part of the Dirac formalism
(e.g. the so called ``complex spectral representation" of Appendix B).
These rules can
only be justified by the full RHS theory.
We shall simply introduce these rules as needed
while referring the reader to the
literature$^{11,14,15)}$ for their justification.

However, since the distinction between the spaces
$\Phi_+$ and $\Phi_-$ will be of physical
importance we need to explain some of their mathematical differences.
This is done using
their mathematical ``realizations".
One often says in mathematics that an abstract (linear
topological) space is ``realized" by a function space if
there exists a correspondence between
each vector $\phi\in\Phi$ and an element (or elements)
of the space of functions (probably in
mathematics functions are more ``real" than vectors).
The basis vector expansions of
Appendix B are examples of mathematical ``realizations".
In this manner the space
$\Phi$ is represented or ``realized" by the standard test function space
(Schwartz space).  The space
$\Phi_+$ (and
$\Phi_-$) is ``realized" by the subspace of Hardy class functions,
where the $+$ ($-$)
refers to analyticity in the upper (lower) half plane of the
second sheet of the complex
energy surface.$^{14,15)}$

In physics the abstact mathematical objects are realized by physical objects.
Thus a
physicists' ``realization" of the linear spaces
$\Phi_+,\ \Phi_-$ and their duals are by quantum
physical objects like states and observables.
The standard quantum theory uses the same Hilbert
space for both states and observables.

In contrast, distinct initial-boundary conditions for state vectors
$\phi^+$ ($e.g.$, in-states
$\phi^+$ of a scattering experiment) and observables
$|\psi^-\rangle\langle\psi^-|$ (e.g., so-called out-states $\psi^-$ of a
scattering experiment) lead to
two different rigged Hilbert spaces$^{12)}$:
\begin{eqnarray}
&\phi^+\in\Phi_-\subset{\cal
H}\subset\Phi^\times_-
&\mbox{for\ in-states\  of\ a\  scattering\
experiment which}\nonumber\\
&&
\mbox{are\ prepared\  by\  a\ preparation\ apparatus,}\nonumber\\
&&\mbox{e.g.\  an\  accelerator}\\
&\psi^-\in\Phi_+\subset{\cal
H}\subset\Phi^\times_+
&\mbox{for\  observables\  or\
out-states, which\  are}\nonumber\\
&&\mbox{measured by a\ registration\  apparatus,}\nonumber\\
&&\mbox{e.g.\  a\
detector}
\end{eqnarray}
The Hilbert space ${\cal H}$ is the same in both RHS's (1) and (2).
However $\Phi_+$ and
$\Phi_-$ are different
but they have more than the zero vector in common,
$\Phi_-\cap\Phi_+\not=\{\emptyset\}$;
in fact $\Phi_-\cap\Phi_+$ is in general infinite dimensional.
To use different mathematical spaces for states
(in-states) and observables (so-called out-``states")
is one of the new features of the RHS
formulation of quantum mechanics.

In (1), $\Phi_-$ describes the possible state vectors experimentally given by
the preparation apparatus (e.g. $\phi^{\rm in}$ or
$\phi^+$ of a scattering experiment) and in (2) $\Phi_+$  describes the
possible observables (e.g., $|\psi^{\rm out}\rangle\langle\psi^{\rm out}|$ or
$|\psi^-\rangle\langle\psi^-|$ of a scattering experiment) experimentally
specified by the detector.
The $|F^\pm\rangle\in\Phi^\times_\mp$ represent quantities connected
with the microphysical
system (e.g. ``scattering states $|E^\pm\rangle$ or decaying states
$|E-i{\Gamma\over2}^\pm\rangle$.
The superscripts for the vectors $\phi^+\in\Phi_-$ and
$\psi^-\in\Phi_+$, and the superscripts in
$|E^\pm\rangle\in\Phi^\times$ for the
eigenkets of the Hamiltonian $H=H_0+H_{int}$
refer to the standard notation of
scattering theory where $\phi^+$ represent
the in-states and $\psi^-$ represent the
out-observables (also called out-states).  The vectors $|E^\pm\rangle$
are related by the
Lippman-Schwinger equations to the eigenkets $|E\rangle$ of $H_0$, and the
superscripts of the eigenkets
$|E-i{\Gamma\over2}^-\rangle\in\Phi^\times_+$ and
$|E+i{\Gamma\over2}^+\rangle\in\Phi^\times_-$ of the
(essentially self-adjoint,
semibounded) Hamiltonian $H$ with complex eigenvalue
($E\mp i{\Gamma\over2})$ are an
extension of the labels in $|E^\mp\rangle$.
(The antithetical subscripts for the spaces have their
origin in the mathematicians notation for
the Hardy class functions.)
In this paper we shall only use the
Gamow vectors
$|E-i{\Gamma\over2}^-\rangle$ which have the property (1.4)
describing exponentially
decaying microphysical objects.
\setcounter{equation}{0}
\section{The Complex 2-dimensional Hamiltonian of the neutral K-system
as a Truncation of the exact Hamiltonian in the RHS.}

The neutral $K$ meson is produced in an inelastic scattering experiment.
We discuss here the case
that this inelastic scattering process produces pure $K^0$ states, as e.g.:
in the reaction:$^{4)}$
\begin{equation}
\pi^- p\to\Lambda K^0\to\Lambda\pi\pi\ .
\end{equation}
The principles of a scattering experiment applied to
this process are depicted in Fig. 1.  The scattering experiment consists of a
preparation part and a registration part.$^{12,15,16)}$  A meson beam
$\pi^-$ (i.e.
in the $II_3y=1,-1,0$ state), which is prepared
as a $\phi^{\rm in}$
before the interaction with the
target
$B$, evolves in the interaction region as a
$\phi^+\in\Phi_-$.  Due to the interaction it makes a transition into the
prepared state of the $K^0$ flavor,
described according to (2.1) by a
$\phi^+_{y=1}\in\Phi_-$.  Thus the
part of the experiment,
which prepares the state $\phi^+_{y=1}$,
consists of the apparatus for the preparation of the
$\pi^-$--beam and the strong interaction with the baryon-target
$B$ (changing the target state from
$p$ to
$\Lambda$).  The registration part
of the experiment
determines the so-called out-``state" $\psi^-$ which is
registered outside the interaction region as e.g. either
$\psi^{\rm out}=|\pi^+\pi^-\rangle$ or $\psi^{\rm out}=|\pi^0\pi^0\rangle$.
Its principal component is the $\pi^+\pi^-$ and/or $\pi^0\pi^0$ detector.
According to (2.2) the out-state, which is actually an
observable $|\psi^-\rangle\langle\psi^-|$,  fulfills
$\psi^-\in\Phi_+$.
In the conventional formulation of scattering theory in the Hilbert
space,
the in-state
$\phi^+$ (and $\phi^{\rm in}$) as well as
the out-observable
$\psi^-$ (and $\psi^{\rm out}$)
can be any vector of the Hilbert space ${\cal H}$.
In reality the $\phi^+$ and $\psi^-$
are subject to different conditions, namely, initial conditions for $\phi^+$
and final conditions for $\psi^-$.  This is described in the new RHS quantum
theory by using different mathematical conditions, $\phi^+\in\Phi_-$
and
$\psi^-\in\Phi_+$,
as shown in
equation (2.1) and (2.2), respectively.

We now consider the state of the meson at times (in the rest frame of the
$K^0$) $t\geq0$.  Here $t=0$ is the time before which the preparation
of the neutral
$K$-meson state (or in general of the $\phi^+$ state) is
completed and after which the registration of
$\psi^{\rm out}\leftarrow\psi^-$ begins.
The time
$t$ is the proper time of the Kaon which in the actual experiment
is measured by the
distance $d$ from the target position (or from the exit
face of the regenerator in
regeneration experiments when
$\phi^+\sim f\phi^+_{y=1}+\bar f\phi^+_{y=-1}$)
to the decay vertex ($t=d m_K/cp$,
where $p$ is the component of the Kaon momentum
along the beam line).
To each decay vertex we associate a microphysical
exponentially decaying
state of the neutral $K$-system,
which is represented by a Gamow vector.
This exponentially decaying Gamow state is a component
of $\phi^+(t)$  that
has dynamically evolved in time from the original,
prepared state
$\phi^+=\phi^+(t=0)$ by the exact Hamiltonian $H$.
(The Hamiltonian $H$
contains all interactions,
including the one responsible for the Kaon decay.)
Since $\phi^+$ is according to (B2)
a superposition of functionals the time evolution
operator is given by the semigroup $e^{iH^\times t}$; thus:$^{17)}$
\begin{equation}
\phi^+(t)=e^{-iH^\times t}\phi^+\phantom{00000}\mbox{or}\phantom{00000}
\phi^+(t)=e^{-i{\IH}^\times t}\phi^+\phantom{00000}t\geq0\ .
\end{equation}

We first want to discuss a theory in which the CP violating
$H_{sw}$ is assumed
to be zero and for which the exact Hamiltonian is denoted by $H$.
Then
[$H$,\ CP]=0 and we choose $H$ and CP as the
complete system of commuting
observables (c.s.c.o.). We ignore all other observables for the
neutral $K$ system except for the
energy operator
$H$, the weak-interaction-free energy operator $H_0$,
the hypercharge $\hat Y$ and the discrete symmetries like CP.
(This means we are
always working in the $K$-meson rest frame where we have
($\vec p_K, j_3,\ j^\pi)=(\vec 0,\ 0,\ 0^-$) which we do not write ).
In place of the
c.s.c.o.
$H$, CP, one can also consider the c.s.c.o.  $H_0,\ \hat Y$.
Besides these two, there is still
another c.s.c.o.:
$H_0$ and CP which we however do not want to consider.
The operators $H$ and
$\hat Y$ do not form a c.s.c.o.

Now we shall make use of some exact results in Appendix B.
We want to contrast the two basis systems
given by the nuclear spectral theorem (B1)
and by the complex basis vector expansion (B2) for the
c.s.c.o. H, CP.  The system of basis vectors is denoted respectively by
\poczn
\begin{equation}
|E,\ cp^-
\rangle\phantom{00000}\mbox{for}
\phantom{00000}0\leq E<\infty,
\end{equation}
and by
\begin{equation}
|z_i,\ cp^-\rangle;\quad\mbox{for}
\quad z_i=z_{K_1},z_{K_2};
\quad\mbox{and}\quad|\omega,
cp;^+\rangle\quad{\rm for}
\qquad
-\infty_{II}<\omega\leq0\ .
\end{equation}
\konn
And we also shall consider the
eigenvectors of the c.s.c.o. $H_0,Y$ which we
denote by
\begin{equation}
|E,y\rangle\qquad{\rm for}\qquad 0\leq E<\infty
\end{equation}
The {\it missing} superscript $^-$ in (4)
indicates that $|E,y\rangle$ are eigenvectors of $H_0$
and not of
$H$  (this latter notation is not in agreement with the
standard notation of scattering theory
where $|E\rangle$ would denote the eigenvector of
$H_{00}= H_0-H_{\rm strong}$).

The vectors
\begin{equation}
|E=m_K,\quad y=\pm1\rangle\qquad {\rm correspond\  to\  the\
usual}\qquad |K^0\rangle,\ |\bar K^0\rangle\ .
\end{equation}
while the vectors
\begin{equation}
|z_1,cp=+,^-\rangle;\quad|z_{2,}
cp=-,^-\rangle\qquad
\mbox{\rm correspond\  to\  the\  usual}\qquad
|K_1\rangle;\enspace |K_2\rangle 
\end{equation}
Therefore we also use the notation:
\addtocounter{equation}{-1}
\poczn
\begin{equation}
|K_{i,}^{\enspace-}\rangle=|z_i,
cp=-(-1)^i,^{\enspace-}\rangle,\quad z_i=m_{K_i}-i
{\gamma_i\over2}\phantom{00000}i=1,2 
\end{equation}
\konn
(up to an arbitrary ``normalization" constant,
cf. (18) below and footnote 20).
This notation also explains why already in equations
(1.3) and (1.4) we employed the
unusual notation
$|K_{1}^{\enspace-}\rangle$ and $|K_{2}^{\enspace-}\rangle$
with a superscript $^-$,  in contrast to the
usual notation $|K_{1}\rangle$ and $|K_2\rangle$.
The $|K_{i}^{\enspace-}\rangle$ are the exact
generalized eigenvectors of the self-adjoint,
semi-bounded Hamiltonian $H$ with complex
eigenvalues $z_i=m_i-i{\gamma_i\over2}$, i.e.
they are the Gamow kets of the neutral
Kaons system. The Gamow kets fulfill {\it exactly }
the (exponential) time evolution equation
(1.4), with $H$ being the infinite dimensional self-adjoint
Hamiltonian operator and not a complex
two dimensional matrix $H^{eff}$.  But (1.4) (and also (1.3))
is to be understood as a functional
equation over the space $\Phi_+$:
\begin{eqnarray}
&\langle e^{iHt}\psi^-|z^-_i\rangle\equiv
\langle\psi^-|e^{-iH^\times t}
|z^-_i\rangle=
e^{-i(m_i-i{\gamma_i\over2})t}
\langle\psi^-|z_i^-\rangle\nonumber\\
&{\rm for\  all}\
\psi^-\in\Phi_+\phantom{00000}
{\rm and}\phantom{00000}t\geq0.
\end{eqnarray}
This means one can only form Dirac bra-ket of (1.4)
with a $\psi^-$ that is an element of the
infinite dimensional space $\Phi_+$ and not in general with
$\phi^+\in\Phi$, cf. footnote 17.

That the vectors $\psi^-$ in (7) can only be of the space $\Phi_+$
($i.e.$ representing
the registered decay products, e.g.
$\psi^{\rm out}=\pi^+\pi^-$) is not a
restriction on physics since we are interested
in decay probabilities or
transition rates into observed out-``states" and
not in arbitrary matrix elements.
That $t\geq0$
in (7), i.e. that the time evolution of the
Gamow vectors is given by a semigroup, not a unitary
group, is derived using the mathematics of the RHS$^{7)}$
and is the  appropriate
restriction for decay processes (arrow of time, microphysical irreversibility).

Using the c.s.c.o. $H, CP$,  we have two choices for the basis vector
expansion:\\[3pt]
1. Dirac basis vector expansion (B1):
\begin{equation}
\phi^+=\sum_{cp=\pm1}\int^\infty_0dE\quad |E,{\rm cp}^+
\rangle b_{\rm cp}(E)
\end{equation}
where the expansion coefficients (energy wave function)
\begin{equation}
b_{cp}(E)=\langle^+E,cp|\phi^+\rangle\in {\cal S}\enspace
\mbox{(Schwartz\ space)}.
\end{equation}
If $\phi^+$ is a state of the neutral Kaon system
then $b_{cp}$(E) are presumed to be peaked
at $E={m_{K_1}}$ for $cp=1$,  and at $E={m_{K_2}}$ for $cp=-$1.  These
functions
may in the
neighborhood of the energy value  $E={m_{K_i}}$
come close to being Breit-Wigner amplitudes (with
widths $\gamma_i$).

The expansion (8) has its
{\it analogue} in Hilbert space quantum mechanics.
Hilbert space quantum mechanics amounts
to the assumption that
the energy wave functions fulfill the conditions
\begin{equation}
b_{cp}(E)\in L^2\qquad {\rm and}\qquad
Eb_{cp}(E)\in L^2\qquad\mbox{(Lebesgue
square integrable)}.
\end{equation}
Then
$b_{cp}(E)$ cannot be a
Breit-Wigner amplitude. Further the assumption (8) with
(9) or (10) causes the well known problems of
deviations from the exponential decay law in
general$^{18)}$ and  some additional problems
specifically for the neutral $K$-system.$^{10)}$

If the prepared state $\phi^+$ is a pure strangeness =$+1$ state
$\phi^+=\phi^+_{y=1}$,
which would usually be denoted by $\phi^+_{y=1}=|K^0\rangle$,
then for that $\phi^+_{y=1}$ the wave functions in (8) would fulfill:$^{19)}$
\begin{equation}
b_{cp=-1}(E)=b_{cp=+1}(E)\equiv{b(E)\over\sqrt{2}}
\end{equation}
This means, a prepared state vector with a definite hypercharge
$y=+1$ or $-1$ has the
Dirac basis vector expansion:
\begin{equation}
\phi^+_{y}=\int^\infty_0dE\ (|E,\
cp=+1,^+\rangle+\mbox{sign}(y)\ |E,cp=-1,^+\rangle)\
{b(E)\over\sqrt{2}}
\end{equation}
If one defines
\begin{equation}
``|K^{0}_{1,2}\rangle"=\int^\infty_0dE|E,cp=+1,-1,^+\rangle b(E)
\end{equation}
and uses the notation
\begin{equation}
\phi^+_{y=1}=|K^{0}\rangle\enspace \qquad  \phi^+_{y=-1}=|\bar
K^{0}\rangle,
\end{equation}
then (12) is the standard expression
\begin{equation}
|K^{0}\rangle={1\over\sqrt{2}}(``|K^{0}_1\rangle"+``|K^{0}_2\rangle")
\enspace\qquad |\bar
K^{0}\rangle=
{1\over\sqrt{2}}
(``|K^{0}_1\rangle-``|K_2^{0}\rangle")
\end{equation}
However
the vectors
`` $|K_{1,2}^{0}\rangle"$
do not give the Lee-Oehme-Yang theory, because they cannot have the
properties (1.4a) and
(1.3a) that one demands of the  usual $|K_{1,2}^{0}\rangle$. Indeed, if one
defines
$|K^{0}_{1,2}\rangle$ by (13), and imposes the precise conditions (10)
given by the
Hilbert space formulation, then one arrives at all kinds of ``CP-violation
problems of the exact
(Hilbert space) quantum theory."$^{10)}$

We shall not use Hilbert space quantum theory
here and thus not make the assumption (10).  Instead we will use
the conditions (2.1), which in terms of the wave function
$b_{cp}(E)=\langle^+E,cp|\phi^+\rangle$ means $b_{cp}(E)\in{\cal
S}\cap{\cal H}_-$ (Hardy class
function from below$^{14)}$).  Then one can also use the Dirac basis vector
expansion
(B1) to obtain (8) for $\phi^+\in\Phi_-$.  But as an alternative to (B1)
one has a second
choice:\\[3pt]
2.  The complex basis vector expansion (B2):
\begin{equation}
\phi^+=|K_1^{0\enspace-}\rangle b_1+|K_2^{0\enspace-}\rangle
b_{-1}+F^-_{cp=+1}+F^-_{cp=-1}
\end{equation}
Here we have defined
\begin{equation}
F^-_{cp}=\int^{-\infty_{II}}_0dE\quad|E,cp^-\rangle b_{cp}(E)
\end{equation}
which we call the background term.  And we have defined the (differently
normalized$^{20)}$) Gamow vectors
\begin{equation}
|K^{0\thinspace-}_i\rangle=|z_i,cp=-(-1)^i,^-\rangle\sqrt{2\pi\gamma_i}\quad
;\quad
z_i=m_i-i{\gamma_i\over2}
\end{equation}
or the microphysical state operator
\begin{equation}
|K_i^{0\enspace-}\rangle\langle^+K^{0}_i|=|z_i,^-\rangle2\pi\gamma_i\langle^
+z_i|
\end{equation}
Since  $|K_i^{0\enspace-}\rangle\in\Phi^\times_+$,
the vectors defined by the r.h.s. of (17)
must also be in
$\Phi^\times_-:\ F^-_{cp}\in\Phi^\times_+$, because
$\phi^+\in\Phi_-$.

According to the theory that underlies (B2), the eigenvalues
\[
z_i=m_i-i{\gamma_i\over2}
\]
in (18) and (19) are the complex energies of $K^{0}_i(i=1,2)$. This means
they are the positions
of the resonance poles for the two resonances $K^{0}_1$ and $K^{0}_2$.
(e.g. of the $S$-matrix
for the scattering process with resonance formation:$^{21)}$
\begin{equation}
\pi\pi\to(K^{0},\bar K^{0})=(K^{0}_1,K_2^{0})\to\pi\pi
\end{equation}
or any other scattering processes in which $K$-resonance formation occurs).
This value is
identical with the generalized eigenvalue of $H$.

The expansion coefficients -- i.e. the coordinates of the vector $\phi^+$
along the basis
vectors (3a) and (3b) -- are according to (B2) given by
\begin{eqnarray}
&&\!\!\!\!\!\!\!\!\!\!\!\!\!\!\!\!\!\!\!\!\!\!\!\!
b_{cp=+1,}=\langle^+K_1^{0}|\phi^+\rangle\quad,\quad
b_{cp=-1}^{\phantom{00000}}=\langle^+K^{0}_2|\phi^+\rangle=
\sqrt{2\pi\gamma_2}\langle^+z_2|\phi^+\rangle\\
&&\!\!\!\!\!\!\!\!\!\!\!\!\!\!\!\!\!\!\!\!\!\!\!\!\!\!\!\!\!\!
b_{cp}(E_{II})=\langle^-E_{II},cp|\phi^+\rangle,\quad
\langle^-E_{II},
cp|\phi^+\rangle=S_{cp}(E_{II})\langle^+E_{II}cp|\phi^+\rangle, 
\end{eqnarray}
where $S(E)$ is the $S$-matrix
analytically continued
to values $E_{II}$ at the negative
real axis on the 2nd sheet (of e.g. the scattering process (20) if we
consider, as we shall,
the decays $K{^0}\to\pi\pi$).

The representation (16) is the special case of (B2) if there are no bound
states (i.e.
no neutral K's with a mass below the $\pi^+\pi^-$ threshold) and if there
are only two
resonance states with $j^\pi=0^-$.  Under these assumptions the representation (16) is
therefore exact (like the spectral theorem) for every $\phi^+\in\Phi_-$.
Whereas the vectors $|K^{0\enspace-}_i\rangle$ are well known basis vectors,
the ``background
integrals" (17) (integrals along the negative real axis of the second sheet)
depend upon the dynamics $H$ or $S$, and upon the prepared state $\phi^+$
and are not well known.$^{23)}$

The best situation is obtained if one can prepare a state $\phi^+$ such
that the background
term is very small.  Then one has from (16) for any pure (i.e. coherent
``mixture" or superposition) neutral Kaon state:
\begin{equation}
\phi^+\approx|K^{0\thinspace-}_1\rangle b_1,+|K^{0\thinspace-}_2\rangle b_{-1}
\end{equation}
This means that for values of
$|b_{cp}(E)|\ll|b_{cp}/\sqrt{\gamma_i}|$
any prepared pure neutral
Kaon state is -approximately- a superposition of
$|K_1^{0\enspace-}\rangle$ and
$|K_2^{0\enspace-}\rangle$.

We now shall take for the prepared state $\phi^+$ a hypercharge eigenstate
as produced e.g.
by the elastic scattering process (1)$^{24)}$
(Gell-Mann and Pais hypothesis) which we call
$\phi^+_{y=+1}$ (corresponding to $|K^{0}\rangle$) or $\phi^+_{y=-1}$
(corresponding
to $|\bar K^{0}\rangle$).   Then
\begin{eqnarray}
&&\phi^+_y=|z_1, cp=+1^-\rangle\tilde b^y_1+|z_2,cp=-1^-\rangle\tilde b^y_{-1}
\nonumber\\
&&+\int^{-\infty_{II}}_0dE\bigl(|E,cp=+1^{\enspace-}\rangle
b^y_1(E)+|E,cp=-1,^-\rangle b^y_{-1}(E)\bigr)
\end{eqnarray}
and using the same arguments that led to
(11),$^{25)}$  one obtains:
\begin{equation}
{1\over\sqrt{2\pi\gamma_1}}
\tilde b^{y=\pm1}_1=\pm b_{-1}^{y=\pm1}\equiv{b\over\sqrt{2}},\qquad
\pm b^{y=\pm1}_1(E)=
b_{-1}^{y=\pm1}(E)\equiv {b(E)\over\sqrt{2}}.
\end{equation}
For a hypercharge eigenstate the approximation (23) then becomes:
\begin{equation}
\phi^+_{y=\pm1}\approx(|K_1^-\rangle\pm|K_2^{\enspace-}\rangle){b\over\sqrt{2}}
\end{equation}
This is the analogue of the standard expressions (15), except that the
$|K_i^{\enspace-}\rangle$
in (26)
are the Gamow vectors with the properly (1.3) that have the exact time
evolution (7) or (1.4).

We now obtain the time evolution (2) of the prepared state $\phi^+_y$ for
$t>0$ using the
representation (24), since for $t\geq0$ the evolution of the Gamow vectors
is known from (7)
\begin{eqnarray}
\phi^+_{y=\pm1}(t)&=&
e^{-iH^\times t}\phi_{y=\pm1}=
{b\over\sqrt{2}}
(e^{-im_1t}e^{-{\gamma_1\over2}t}|K_1^{\enspace-}\rangle\pm
e^{-im_2t}e^{-{\gamma_2\over2}t}|K_2^{\enspace-}\rangle)\nonumber\\
&\!\!\!\!\!+&\!\!\!\!\!\int^{-\infty_{II}}_0
dE\bigl(\pm|E,cp=+1^-\rangle+|E,cp=
-1^-\rangle\bigr)e^{-iEt}{b(E)\over\sqrt{2}},
\end{eqnarray}
Note that according to (27), $e^{-iH^\times t}$ cannot transform from the
background term
$(F_{+1}^{\enspace-}+F^{\enspace-}_{-1})$
to
$|K^{\enspace-}_1\rangle$ or
$|K^{\enspace-}_2\rangle$ and it cannot transform
from
$|K_2^{\enspace-}\rangle$ to $|K_1^{\enspace-}\rangle$.  This means our
theory does not predict
anything that could be interpreted as  ``vacuum regeneration of $K_1$ (or
$K_S$) from
$K_L$."$^{10)}$

The approximation of a prepared state for very small background
term,
$|b(E)\sqrt{\gamma_{i}}/b|\ll1$, given by (26), has the time evolution:
\begin{equation}
\phi^+_{y=\pm}(t)\equiv
e^{-iH^\times
t}\phi^+_{y=\pm1}={b\over\sqrt{2}}(e^{-im_1t}e^{-{\gamma_1\over2}t}|K_1^-
\rangle\pm
e^{-im_2t}e^{-{\gamma_2\over2}t}|K_2^{\enspace-}\rangle)
\end{equation}
This means that the RHS quantum theory reproduces to a certain extent the
Lee-Oehme-Yang theory, and if $\phi^+$ can be prepared such that the
background term
$|b_{cp}(E_{II})|$ is negligibly small, then the Lee-Oehme-Yang theory
emerges as its
approximation.  Vice versa, the worthiness of the Lee-Oehme-Yang theory can
be taken as a
measure of how small the background term
$F_{cp}^{\enspace-}\in\Phi^\times_+$ in (16) and (24) can be made if the
$K^0$-state
$\phi^+_y\in\Phi_-$ is suitably prepared.

Thus the ``complex spectral" resolution (B2) of the RHS formulation of
quantum mechanics chooses
the basis system in $\Phi^\times_+$ such that the two dimensional space
${\cal H}_2$ of the
standard neutral Kaon model is spanned by the generalized basis vectors $|K_1^{\enspace-}\rangle,
|K_2^{\enspace-}\rangle\in\Phi^\times_+$.  One does not have to make any
special assumptions
about a complex effective Hamiltonian.  The Hamiltonian $H$ (precisely its
closure $H^\dagger$)
is just required to have the standard properties:  It is a self-adjoint
operator, bounded from
below.

The two dimensional matrix $H^{eff}$ emerges as the matrix of this infinite
dimensional
operator $H$ in the two dimensional subspace ${\cal
H}_2\subset\Phi^\times_+$.  Thus the RHS
formulation justifies the effective Lee-Oehme-Yang theory -- in contrast to
the exact Hilbert space
formulation, which contradicts it.$^{10)}$  The RHS formulation also gives
some additional
results, such as the semigroup $(t\geq0)$ time evolution of the Kaon decay,
(27), (28).
Furthermore,
for the
Kaon system in particular, it predicts the background term, which, however
small, must be
different from zero (because $\phi^+\in\Phi_-$ cannot be a superposiiton of
two vector
$|K_1^{\enspace-}\rangle, |K_2^{\enspace-}\rangle\in\Phi^\times_+$).

The existence of this background term in (27) has some significant
consequences, which may or
may not be of practical importance since $|{\sqrt{\gamma} b(E)\over b}|\ll1$ 
could be too
small to be observed.
However, since in the case of CP violation and direct CP-violation one is
talking of effects of
the order of $10^{-3}$ or 10$^{-6}$ and since one also discusses CPT
violations and violations of
microphysical quantum coherence,$^{26,5)}$ which are orders of magnitude
smaller
than $10^{-3}$,
a
discussion of possible effects from the background terms is warranted.
\setcounter{equation}{0}
\section{Long-time $2\pi$ decays of neutral $K$ without CP
violating Hamiltonian}

The quantity that is measured in the neutral $K$ experiments$^{27)}$ is the
instantaneous decay
rate of the $K^{0}$'s  into $\pi\pi$, ($\pi^+\pi^-$ or
$\pi^0\pi^0$).  In analogy to the instantaneous transition rate of
$\phi^{eff}(t)$ into
$\pi\pi$, which is given by (1.8) with (1.7) and (1.5), we take for the
instantaneous transition
rate of the prepared state $\phi^+(t)$ into $\pi\pi$ the
matrix element
${\hbox{$<\pi\pi|H_w|\phi_{y=1}^+(t)\rangle$}}$,
where
$H_w$ is the CP conserving interaction Hamiltonian.
The difference
between the standard phenomenology and the model that we want to consider
in this section
is the following:  Our Hamiltonian is
\begin{equation}
H=H_0+H_w\qquad{\rm with}\qquad[H,CP]=0\qquad[H_w,CP]=0
\end{equation}
and our $K^0$ state ($\phi^+_{y=1}$ prepared in the inelastic scattering
process (3.1)) is
given by (3.27), which evolves with a preferred time direction, $t\geq0$.
In the effective
theory summarized in section 1, the Hamiltonian is $\IH$ given by (1.2),
but the time evolving
state vector  is given by (1.6) (and though it is only used for the forward
time direction there
is no theoretical reason that $\phi^{eff}(t)$ could not also evolve
backward in time).  For our
model we define the ratio
$R(t)$ by replacing in (1.8)
$\IH_{in}$ by $H_w,\ \phi^{eff}$ by $\phi^+$ and
$<\pi\pi|\IH_{int}|K_S\rangle$ by $<\pi\pi|H_w|K_1^{\enspace-}\rangle$, since
$|K_S^{\enspace-}\rangle\to|K_1^{\enspace-}\rangle$ for an $H$ given by
(1).  Thus the
normalized instantaneous
rate as a function of $t$ (proper
time in the $K^0$ rest frame) is given by:$^{28)}$
\begin{equation}
R(t)=\Big|{<\pi\pi|H_w\big|
\phi_{y=+1}^{+}(t)>\over<\pi\pi|H_w|K_1^{\enspace-}>}\Big|^2
\end{equation}
We calculate $R(t)$ using our theoretical result (3.27) and obtain by
(inserting (3.27) into
(2)):
\begin{equation}
R(t)=\Big|e^{{-\gamma_1\over2}t}
e^{-im_1t}{b\over\sqrt{2}}+
\int^{-\infty_{II}}_0d
E{<\pi\pi|H_w|E,cp=+1^{\enspace-}>\over<\pi\pi|H_w|K_1^{\enspace-}>}
{b(E)\over\sqrt{2}}
e^{-iEt}\Big|^2,
\end{equation}
since $<\pi\pi|H_w|K_2^{\enspace-}\rangle=0$, and
$<\pi\pi|H_w|E,cp=-1,^{\phantom{0}-}\rangle=0$  due to (1).
In the same way as for the
$R(t)$ of
(1.9), the first term vanishes for large values of
$t\approx20{1\over\gamma_s}$ ($e^{-{1\over2}\gamma_st}\approx10^{-5})$ so that
\begin{eqnarray}
&&R(t=20\tau_s)
=\Big|{1\over\sqrt{2}<\pi\pi|H_w|
K_1^{\enspace-}>}\Big|\nonumber\\
&&\times\int_0^{-\infty_{II}}
dE<\pi\pi|H_w|E, cp=+1^{\enspace-}>b(E)e^{-iEt}\Big|^2
\end{eqnarray}

Theoretically, not much can be said at this stage about the integral on the
r.h.s. of (4).
We write it using (3.22) as:
\begin{eqnarray}
&&\hspace{-25pt}I(t)\nonumber\\
&&\hspace{-25pt}=-\int^0_{-\infty_{II}}
dE<\pi\pi|H_w|E,cp=+1^{-}>
S_{cp=+1}(E)<^+E,cp=+1|\phi^+>e^{-iEt}\nonumber\\
&&\hspace{-25pt}=-\int^0_{-\infty_{II}}
dE<\pi\pi|H_w|E,cp=+1^+><^+E,cp=+1|\phi^+>e^{-iEt}
\end{eqnarray}
Then (3) can be written as
\begin{equation}
R(t)=\Big|e^{-{\gamma_1\over2}t}e^{-im_1t}{b\over\sqrt{2}}+
{I(t)\over<\pi\pi|H_w|K_1^{\enspace-}>\sqrt{2}}\Big|^2,
\end{equation}
The integral $I(t)$ describes that part of the transition
$\phi^+\to\pi\pi$ which does not go through $K_1^{\enspace-}$ resonance
formation.
We know that $<^+E|\phi^+\rangle\in S\cap{\cal H}_-$ (Hardy class function
of the lower half plane,
second sheet).
If we also knew that $\langle\pi\pi|H_w|E^-\rangle S(E)\in S\cap{\cal
H}_-$, then one could
prove (using the Riemann-Lebesgue lemma for Hardy class functions)
that
$I(t)$
also
decreases
for increasing $t$, but it decreases
less than an
exponential $e^{-\gamma t}$.
Even though we do not have this information
let us
assume that the non-resonant background $I(t)$ will survive the exponential
$e^{-\gamma_1t}$:
\begin{equation}
|I(t)|^2\geq\  {\rm const.}\  e^{-\gamma_1t}\qquad{\rm for}\qquad
t\geq20{1\over\gamma_1}\ .
\end{equation}
The magnitude of
$\langle^+E_{II},cp=1|\phi^+\rangle$
(which can be calculated from \mbox{$\langle^+E|\phi^+\rangle$} 
on the positive real
axis i.e. at physical
values of $E$ because of its Hardy class property$^{14)}$)
depends upon $\phi^+$ i.e. upon the preparation of the $K^0$
state.
Since experimentally the $\pi^+\pi^-$ are selected
such that their invariant mass is near the center of a
$K^0$-Breit-Wigner energy distribution, the magnitude of the non-resonance
contributions
\mbox{$|\langle^+E_{II}|\phi^+\rangle|$} to the observed $R(t)$ will be small.  
But we need
only a small
contribution
on the r.h.s. of (4) in order
to explain the experimental value (1.13) for $R(20\tau_s)$.  Mathematically
\mbox{$\langle^+E_{II}|\phi^+\rangle$} must be different from zero, 
because \mbox{$\phi^+\in\Phi_-$}.
But its magnitude
could be arbitrarily small, too small to account even for the small value
(1.13)
of
$R(20\tau_1)$.  Thus the question is not whether such a term
\mbox{$|I(t)/\langle\pi\pi(H_w|K^{\enspace-}_1\rangle|$}, which decreases 
more slowly in time
than the
exponential $e^{{-\gamma_1}t}$,
exists
but
whether this term in (6) has
the right magnitude to explain the value (1.13).

Since we
have no theoretical prediction, we shall use the second term on the r.h.s.
of (6) as
phenomenological parameter.

The phenomenological value that we obtain for
this background integral from the values (1.12),$^{29)}$ and
(1.13) is:
\begin{equation}
\Big|{I(t=20\tau_1)\over\langle\pi\pi|H_w|K_1^{\enspace-}\rangle}\Big|
\approx2.23\cdot10
^{-3}\ .
\end{equation}
This means that if the non-resonant contributions to the transitions of the
prepared state
$\phi^+$ into $\pi\pi$ are about 2.23$\cdot10^{-3}$ of the $K_1$-resonance
term, then these
contributions can explain the Princeton effect without the assumption of a
CP-violating
Hamiltonian.  Theoretically 
$|{I(t)\over\langle\pi\pi|H_w|K_1^{\enspace-}\rangle}|$
could of course be much
smaller than
$10^{-3}$, in fact it can be arbitrarily small as long as it is not equal
to zero and still
fulfill the mathematical conditions that $\phi^+\in\Phi_-$.

To display the distinction between the way the effective Lee-Oehme-Yang
theory explains the
existence of a long time $K^0\to\pi\pi$ decay mode and the way the exact
theory in the RHS
explain this effect, we compare the state vector for $K$-short and $K$-long
in both theories.
According to (1.1) we have (except for some normalization):
\poczn
\begin{eqnarray}
|K_S^{\enspace-}>&=&|K_1>+\epsilon_S|K_2>\\
|K_L^{\enspace-}>&=&\epsilon_L|K_1>+|K_2>
\end{eqnarray}
\konn
Here $|K_S^{\enspace-}>$ and $|K_L^{\enspace-}>$, and not
$|K_1>$ and
$|K_2>$ are assumed to be the Gamow
vectors,  and $\epsilon_L\approx\epsilon_S\approx\epsilon\approx10^{-3}$
are very small.  In
analogy to this we define
\poczn
\begin{eqnarray}
\phi^+_\sigma&=&|K_1^{\enspace-}>+{1\over b}F_{cp=-1}^{-}\\
\phi^+_\lambda&=&{1\over b}F^-_{cp=1}+|K_2^{\phantom{0}-}>
\end{eqnarray}
\konn
where $|K_{1,2}^{\phantom{000}-}>$
are the Gamow vectors and $F_{cp}^{\enspace-}$ are the vectors (3.17) in
the expansion (3.16).  Then
$\phi^+_y$ of (3.24), (3.25) can be written as
\begin{equation}
\phi^+_{y=\pm1}=
\phi^+_\sigma{b\over\sqrt{2}}
\pm\phi_\lambda^+{b\over\sqrt{2}}\ .
\end{equation}
This formula is the analogue of (1.5) (with $a_S=\pm a_L={b\over\sqrt{2}}$)
of the effective
theory.

Comparing (10) with (11) we see that
\begin{center}
\begin{minipage}{0.8\textwidth}
$\phi^+_\sigma$ are, like the $|K_S>$, 
mostly
$cp=+1$ vectors with  a  small $cp=-1$ contribution.
\end{minipage}\\
\begin{minipage}{0.8\textwidth}
$\phi^+_\lambda$ are, like the $|K_L>$, 
mostly
$cp=-1$ vectors with  a  small $cp=+1$ contribution.
\end{minipage}
\end{center}
But whereas in (9) the small admixtures of the opposite parity is given by
$\epsilon|K_2>\in{\cal H}_2$ and
$\epsilon|K_1>\in{\cal H}_2$ respectively, the opposite parity admixtures
in (10),
${1\over b}F^-_{cp=-1}$ and
${1\over b} F^-_{cp=+1}$ respectively, are not
elements of ${\cal H}_2$.  (Recall that ${\cal H}_2$
$\subset\Phi^\times_+$ is the space spanned by the vectors
$|K_1^{\enspace-}>,|K_2^{\enspace-}>)$.  Since in the effective theory one
has only the space
${\cal H}_2$, one had to postulate an effective Hamiltonian $\IH$ with
$[\IH, CP]\not=0$ in order to
obtain (9b) and
\begin{equation}
<\pi\pi|\IH|K_L^{\enspace-}>=
\epsilon<\pi\pi|\IH|K_1^{\enspace-}>+<\pi\pi|\IH|K_2^{\enspace-}>\not=0
\end{equation}
In the infinite dimensional RHS formulation one has other vectors outside
of ${\cal H}_2$ which have
the opposite parity so that
\begin{equation}
<\pi\pi|H|\phi_\lambda^+>={1\over b}<\pi\pi|H|F^-_{cp=+1}>\not=0
\end{equation}
even though $[H,CP]=0$.  This term gives the $\pi\pi$ transitions for large
$t\approx20\gamma_s$
when the amplitude $<\pi\pi|H|\phi^+_\sigma>$ is suppressed by a factor of
$e^{-\gamma_1t/2}$.  Thus the analogue of
$\epsilon<\pi\pi|\IH|K_1^{\enspace-}>$ is in the RHS
formulation given by ${1\over b}<\pi\pi|H|F^{\enspace-}_{cp=+1}>$ which has
nothing to do with CP
violation.
In the exact RHS theory with non-CP-violating H there can, however, not be
an anologue of the
direct CP-violation amplitude $<\pi\pi|\IH|K_2^{\enspace-}>$.

The theoretical CP problems that the Princeton effect caused
for the Lee-Oehme-Yang theory
clearly had its origin in the confinement of the theory to the
two-dimensional space ${\cal
H}_2$.  This confinement to ${\cal H}_2$ is an ad hoc assumption which
cannot be justified in an
infinite dimensional space of states, except as the zeroth order of
perturbation theory:   If
one considers
$H_0=H-H_w$ as
the exact Hamiltonian and $H_w$ as
a perturbation of $H$ and if one takes the eigenvectors $|K_1^{\enspace-}>$ and
$|K_2^{\enspace-}>$ of $H$ as the zeroth order eigenvectors of $H_0$, then
one knows from
perturbation theory that the higher order eigenvectors of $H_0$ are
generally not linear
combination of the zeroth order eigenvectors.  The exact eigenvectors
$|K^0>, |\bar K^{0}>$
of $H_0$ are
then given by
the highest ($\infty$) order of perturbation theory.  They
therefore cannot in general be linear combinations of the
zeroth eigenvectors
$|K_1^{\enspace-}>$ and
$|K_2^{\enspace-}>$ only.  But this is exactly what would be required in
(1.5) (for
$a_S=a_L={1\over\sqrt{2}}$ or
$\phi^{eff}=|K^0>$).  Though the complex basis vector expansion (B2) was
the origin of the idea
that the Princeton effect can be explained without a CP-violating
Hamiltonian, it
is not really needed in order to ask the question why the prepared state
$|K^{0}>$ should be
expandable (with
an accuracy of $10^{-3}$)
only in terms of vectors of the two-dimensional ${\cal H}_2$.
The existence of a finite ($>2$) or infinite number of linearly independent
$(\hbox{\rm basis-)}$ vectors -- irrespective of what definition of
convergence one uses -- is
already sufficient to see the problem.  The RHS is needed to
explain the existence of the eigenvectors (1.3a) with the property (1.4a) for $t\geq0$, and
to justify the inclusion of  these eigenvectors in a
complete basis system for the prepared state vectors
$\phi^+_{y=1}\in\Phi_-$,
(where
$\phi^+_{y=1}$ corresponds to $|K^0>$).
For these properties one needs the mathematical (topological) completion of
the linear
scalar product space $\Psi$ (Appendix A).
\setcounter{equation}{0}
\section{The Standard Phenomenological Description of CP Violation
as a Truncation of the
Exact Theory in the RHS}

Even if the $\pi\pi$ decay of the long lived neutral $K$ meson state can be
explained without a CP
violating interaction Hamiltonian, it does not mean that there is no CP
violation of the
Hamiltonian.  As was already mentioned above, the background term in (3.24)
-- though it needs to
be there -- may be so small ($|\sqrt{\gamma_i}b(E)/b|\ll10^{-3}$) that it cannot
account for the
Princeton effect.  Also, the observed time dependence of the transition
amplitude (1.8), in
particular the interference term in (1.9), may be such that it cannot be
explained by the
background integrals in (4.3) or (3.27).
Therefore, we want to
apply now the same exact theory of sections 2 and 3 to the
CP-violating Hamiltonian
\[
\IH=H+H_{sw}=H_0+H_w+H_{sw}\quad;\quad[\IH,\ CP]\not=0 \ .
\]
Then, in place of the eigenvectors $|K^-_1\rangle,|K^-_2\rangle$ of $H$, we
use the
eigenvectors
$|K^-_S\rangle,|K_L^-\rangle$ of $\IH$ in the complex basis vector
expansion (B2) of the K-meson
state vector $\phi^+_y$:
\begin{eqnarray}
\phi^+&=&{1\over\sqrt{2}}(|K_S^{\enspace-}>b_S+|K_L^{\enspace-}>b_L)
\nonumber\\
&+&\sum_\beta\int^{-\infty_{II}}_0dE|E,\beta^->b_\beta(E)/\sqrt{2}\ .
\end{eqnarray}
In here, $|E,\beta^->$ are the generalized eigenvectors of $\IH$, and
$\beta$ are the degeneracy
quantum numbers (where now $\beta\not= cp$).  The complex basis vector
expansion (1) is, as
before, very general and exact, under the assumption that
$|K_S^{\enspace-}>$ and
$|K_L^{\enspace-}>$ are the only
Gamow vectors with the right quantum numbers for the neutral $K$-system.
The complex
expansion coefficients $b_S, b_L$ and
$b_\beta(E)$
$=<E,\beta|\phi^+>$
depend again predominantly upon the experimental conditions for the
preparation of
$\phi^+$.
Again for mathematical reasons
$b_\beta(E)\in{\cal H}_-^2\cap S$ cannot be exactly zero but it could be
arbitrarily small
(mathematics provides no information about the order of magnitude
involved).  Thus if
$|{\sqrt{\gamma_i}b(E)\over b_{S,L}}|\ll10^{-3}$, then, as far as the Princeton
effect (which is of
order
$10^{-3}$) is concerned, every pure neutral-$K$ state obtained from (1) is
adequately approximated
as
\begin{equation}
\phi^+\approx(|K_S^{\enspace-}>b_S+|K_L^{\enspace-}>b_L)
{1\over\sqrt{2}}\sim|K_S^{\enspace-}>\rho+|K_L^{\enspace-}>\ .
\end{equation}
Since the time evolution of the Gamow vectors $|K_S^{\enspace-}>$ and
$|K_L^{\enspace-}>$ is derived (not assumed) to be given by (1.4b) (with
the additional result
that
$t\geq0$), we have obtained
in the RHS an exact theory which contains
the standard phenomenological description of the neutral $K$-system with
CP-violation as an approximation.  No new physics has been developed,
but the standard phenomenological description has been given an exact
theoretical foundation.
The background terms in (1) may play (an observable) role in other experimental
investigations.$^{26)}$

Since the basis vectors on the r.h.s. of (1) are generalized eigenvectors
of ${\IH}$, the
time evolution operator $e^{-i\IH^\times t}$ is diagonal:
\begin{eqnarray}
\phi^+(t)&=&e^{-i\IH^\times t}\phi^+={1\over\sqrt{2}}
(e^{-im_St}e^{-{\gamma_S\over2}t}|K_S^{\enspace-}>b_S+e^{-im_Lt}e^{-{\gamma_
L\over2}t}|K_L^{\enspace-}>b_L)\nonumber\\
&+&\sum_\beta\int^{-\infty_{II}}_0dEe^{-iEt}|E,\beta^{\enspace-}>b^\beta(E)/
\sqrt{2};\phantom{0000}
t\geq0
\end{eqnarray}
In particular $|K_L^{\enspace-}>$ cannot evolve by its own Hamiltonian
$\IH$ (i.e. without
additional interaction with a regenerator) into $|K_S^{\enspace-}>$ or
vice-versa (i.e.
there is no ``vacuum regeneration of $K_S$ from $K_L$"), and neither can $K_S$ be regenerated
due to
$e^{-i\IH t}$ from the background term
\[
F_\beta^-=\int^{-\infty_{II}}_0dE|E,\beta^{\enspace-}>b^\beta(E)/\sqrt{2}\ .
\]
The Gamow vectors $|K_S^{\enspace-}>$ and $|K_L^{\enspace-}>$  evolve
(as a consequence of their definition from the resonance poles)
irreversibly
and obey the exact exponential decay law (1.4b):
\begin{eqnarray}
e^{-i\IH^\times
t}|K_S^{\enspace-}>&=&e^{-im_St}e^{-{\gamma_S\over2}t}|K_S^{\enspace-}>\qquad
,\qquad t\geq0\nonumber\\
e^{-i\IH^\times
t}|K_L^{\enspace-}>&=&e^{-im_Lt}e^{-{\gamma_L\over2}t}|K_L^{\enspace-}>
\qquad,\qquad t\geq0
\end{eqnarray}
There is no additional term on the r.h.s. of (4), in contrast to exact
infinite dimensional
theories in the Hilbert space.$^{10)}$  Also, $|K_S^{\enspace-}>$ and
$|K_L^{\enspace-}>$ cannot be
expressed in terms of
$\phi^+_{y=+1}$ and
$\phi^+_{y=-1}$, or any other finite or infinite
superposition of
$\phi^+_y\in\Phi_-$, due to the terms $F_\beta^-\in\Phi^\times_+$ on the
r.h.s. of (1).  The
time evolution in (3) is irreversible,
$t\geq0$, and $\phi^+(t)\in\Phi^\times_+$.  This means that it can
only be evaluated as a functional $<\psi^-|\phi^+(t)>$ at $\psi^-\in\Phi_+$
(which
represent  observed decay products
like $\psi^-=|\pi\pi>$).  In particular the
functional $\phi^+(t)$ cannot be evaluated
at $\phi^+\in\Phi_-$ i.e. the quantity
$\hbox{$<\phi^+_{y^{\prime}}|\phi^+_y(t)>$}$=\break\hfil
$<\phi^+_{y'}|e^{-iIH^\times
t}|\phi^+_y>$,  which would represent a vacuum regeneration
amplitude of $\bar K^0(y^{\prime}=-1)$ from
$K^0(y=+1)$, and vice versa, makes no sense in our theory.  These kind of
quantities have also no
observable meaning since no experiment can measure the probability for a
``transition" from
an in-state
$\phi^+_{y=1}$ into another in-state
$\phi^+_{y=-1}$.
\setcounter{equation}{0}
\section{Summary and Conclusions}

The purpose of this paper was to use the neutral $K$-system of two
interfering resonances to
test some aspects of the RHS-quantum theory of microphysical
irreversibility.  We limited our
investigation to the hypothesis that the two decaying $K$-states are ordinary,
first order, S-matrix-pole resonances (by the choice of $(B2)$) since the
standard theory with
complex effective Hamiltonian makes the same hypothesis (by the choice of a
diagonalizable complex Hamiltonian matrix rather than a Jordan block).
Then we saw that the effective Lee-Oehne-Yang theory is a subtheory of the
exact theory in
the RHS.
It must be emphasized that this is not the case for the exact theory in
Hilbert space
because the Hilbert space theory
does not allow for a complex basis vector expansion.  As a bonus, we saw
that the
remainder of the
exact theory, which is always ignored in the two dimensional effective
subtheory, leads to a non-zero $2\pi$ decay rate of the neutral $K$-system for
large time even if we choose a CP conserving Hamiltonian.  This may or may
not be of practical
significance since at this stage nothing can be said about its magnitude.

Many more experimental properties are known about the instantaneous
transition rate
$|<\pi\pi|H_{int}|\phi^+(t)>|^2$ than have been used in our discussion in
this paper.$^{27)}$
To
make adequate use of these properties the background terms
$<\pi\pi|H_w|F^{\enspace-}_{cp}(t)>$ need to be investigated further
and more of its characteristics needs to be known than just the property that it
decreases slower than exponentially in time.  Of particular interest is the
transition rate at
instances around
$t=12\tau_s$ where it has been fitted$^{27)}$ to the interference term
$\cos(\Delta
mt+\varphi)$ of (1.9); a result which, in that form, can probably not be
obtained from the
background term.  These questions will have to be discussed in a subsequent
paper.
\appendix
\setcounter{equation}{0}
\section{From a Pre-Hilbert Space to a Rigged Hil\-bert Space}

A pre-Hilbert space is a linear space $\Psi$ with a scalar product.
This scalar product is
denoted
by
\begin{equation}
(\psi,\ F)\quad{\rm or\  by}\quad\langle\psi|F\rangle\ .
\end{equation}
The pre-Hilbert space is without any topological structure; that means
neighborhoods, the convergence of infinite sequences, topological
completeness, continuous operators,
continuous functionals,
dense subspaces, etc. are not defined.  This
space is what physicists mostly use for their calculations,  (together with
a few
additional rules), when they speak of the Hilbert space.

The Hilbert space ${\cal H}$ of mathematicians is a much more complicated
structure. In order to make it topologically complete,  its elements are not
represented by functions (wave functions), but by classes of functions
whose elements differ on a set of Lebesgue measure zero, a mathematically
complicated and physically useless concept (because the apparatus resolution is
described by a smooth function, not a set of Lebesque square integrable
functions).  The RHS is the same linear space
$\Psi$ only with different topological completions:  one completes
$\Psi$ with respect to a topology that is stronger than the topology given by
the Hilbert-space norm (e.g., one uses a countable number of norms) to obtain
the
space $\Phi\subset{\cal H}$ and considers in addition the topological
dual to $\Phi$ i.e., the space of {\it continuous} antilinear functionals of
$\Phi$ denoted by $\Phi^\times$.  Then one obtains the triplet of completions of
$\Psi$ (all differing from $\Psi$ only by limit elements), the Gelfand triplet
or Rigged Hilbert-space:
\begin{eqnarray}
\Phi\subset{\cal H}={\cal H}^\times\subset\Phi^\times
\hphantom{000000}&&\\
\mbox{with  elements ``bra" and ``ket"}\phantom{0000}
\langle\phi|\in\Phi\phantom{000000000}|F\rangle\in\Phi^\times&&\nonumber\\
\mbox{or ``ket" and `` bra"}\phantom{0000}
|\phi\rangle\in\Phi\phantom{000000000}\langle F|\in\Phi^\times&&
\end{eqnarray}
One widespread example for $\Phi$ is the Schwartz space $\mathcal{S}$ 
(i.e. $\Phi$ is
often
``realized" by the space of functions $\mathcal{S}$).

The vectors $\phi\in\Phi$ (in their form as either kets $|\phi\rangle$ or bras
$\langle\phi|)$ represent physical quantities connected with the experimental
apparatuses (e.g. a state $\phi$ defined by a preparation apparatus or an
observable $|\psi\rangle\langle\psi|$ defined by a registration apparatus
(detector) fulfills $\phi,\psi\in\Phi)$, the vectors $\langle F|$ or
$|F\rangle\in\Phi^\times$ represent quantities connected with the microphysical
system (e.g. ``scattering states" $|E\rangle$ or decaying states
$|E-i\Gamma/2\rangle)$.

A general observable is now represented by a bounded operator $A$ in $\Phi$ (but
in general by an unbounded operator $\bar A$ or $A^\dagger$ in ${\cal H}$) and
corresponding to the triplet (A2) one has now a triplet of operators
\begin{equation}
A^\dagger|_{\Phi}\subset A^\dagger\subset A^\times
\end{equation}
In here $A^\dagger$ is the Hilbert space adjoint of $A$ (if $A$ is essentially
self adjoining then $A^\dagger=\bar A$), $A^\dagger|_{\Phi}$ is its
restriction to the space $\Phi$, and the operator $A^\times$ in $\Phi^\times$ is
the conjugate operator of $A$ defined by
\begin{equation}
\langle A\phi|F\rangle=\langle\phi|A^\times F\rangle\qquad
{\rm for\ all}\quad\phi\in\Phi\quad
{\rm and\  all}\quad
|F\rangle\in\Phi^\times.
\end{equation}
By this definition, $A^\times$ is the extension of the operator $A^\dagger$ to
the space $\Phi^\times$ (and not the extension of the operator $A$ which is most
often used in mathematics).  A very important point is that the operator
$A^\times$ is only defined for an operator $A$ which is continuous
(and bounded)
in
$\Phi$, then $A^\times$ is a continuous (but not bounded) operator in
$\Phi^\times$.  It is impossible in quantum mechanics (empirically) to restrict
oneself to continuous (and therefore bounded) operators $\bar A$ in ${\cal H}$.
But one can restrict oneself to algebras of observables $\{A,B\ldots\}$,
described by continuous operators in
$\Phi$,
if the topology of $\Phi$ is suitably chosen.  Then $A^\times,\ B^\times\ldots$
are defined and continuous in $\Phi^\times$.  If
$A$ in (5) is not self-adjoint then
$A^\dagger|_\Phi$ need not be a continuous operator in $\Phi$ even if $A$ is,
but one can still define the conjugate $A^\times$ which is continuous in
$\Phi^\times$.

A generalized eigenvector $F\in\Phi^\times$ of an operator $A$ is defined by
\begin{equation}
\langle A\phi|F\rangle=\langle\phi|A^\times
F\rangle=\omega\langle\phi|F\rangle\quad
{\rm for\ all}\quad\phi\in\phi
\end{equation}
where the complex number $\omega$ is called the generalized eigenvalue.  This
is also written as
\begin{equation}
A^\times|F\rangle=\omega|F\rangle\ .
\end{equation}
For an essentially self-adjoint operator ($A^\dagger=\bar A$ = closure of $A$)
this is often also written
(following Dirac)
as
\begin{equation}
A|F\rangle=\omega|F\rangle\ ,
\end{equation}
especially if one suppresses the mathematical subtleties and acts as if one has
just a linear scalar-product space $\Psi$.  The generalized eigenvalues
$\omega$ for self-adjoint operators $A^\dagger$ need not be real.
\setcounter{equation}{0}
\section{The complex basis vector expansion}

The most important result of the new mathematical theory of quantum physics in
the rigged Hilbert space is the complex eigenvector expansion.  This is the
generalization of the elementary basis vector expansion of a 3-dimensional
vector,
\[
{\bf x}=\sum_{i=1,2,3}{\bf e}_i({\bf e}_i\cdot {\bf x})=\sum {\bf e}_ix_i,
\]
to
the expansion of vectors $\phi^+\in\Phi_-$ using as basis vectors the
generalized eigenvectors $|z^-_{R_i}\rangle$ of self-adjoint operators $H$ with
complex eigenvalues
$z_{R_i}$ and $z$.

Earlier developments towards this generalization were the fundamental theorem of
linear algebra which states that for every self-adjoint operator $H$ in a
$n$-dimensional Euclidian space ${\cal H}_n$ there exists an orthonormal basis
$e_i\ldots e_n$ in ${\cal H}_n$ of eigenvectors $He_i=E_ie_i$,  i.e.,
$f\in{\cal H}_n$ can be written as $f=\sum^n_{i=1}e_i$ ($e_i,f)$.  This theorem
generalizes to the infinite dimensional Hilbert space ${\cal H}$, but only for
self-adjoint operators $H$ which are completely continuous (also called compact
operators which include Hilbert-Schmidt, nuclear, traceclass operators).  For an
arbitary self-adjoint operators  one has to go outside the space to find a
complete basis system of eigenvectors (generalized).

The {\it first} step in this direction is the Dirac basis vector expansion
which in mathematical terms is called the nuclear spectral theorem.  It states
that for every $\phi\in\Phi$
\begin{equation}
\phi=\int^{+\infty}_{0}dE|E^+\rangle\langle^+E|\phi^+\rangle
\quad+\quad\sum_n|E_n)(E_n|\phi)\quad
{\rm for}\quad
\phi\in\Phi
\end{equation}
In here, $|E_n$) are the discrete eigenvectors of the exact Hamiltonian
$H=K+V$, (describing the bound states) $H|E_n)=E_n|E_n)$, and $|E^+\rangle$ are
the generalized eigenvectors (Dirac kets) of H fulfilling $\langle
H\chi|E^+\rangle=\langle\chi|H^\times|E^+\rangle=E\langle\chi|E^+\rangle$ for
all
$\chi\in\Phi$, cf. (A6).
The ``coordinates" of the vector $\phi$ with respect to the continuous basis
$|E^+\rangle$, i.e. the set of energy wave functions $\langle^+E|\phi^+\rangle$
form a ``realization" of the space $\Phi$ by a space of functions.  We call
$\phi\in\Phi$ well-behaved" if
$\langle^+E|\phi^+\rangle$ is a well-behaved function, i.e. of the Schwartz
space $\mathcal{S}$.
The $|E^+\rangle$ correspond to the contnuous
spectrum (describing scattering states) and the integration extends over the
spectrum of $H:0\leq E<\infty$.  In place of the $|E^+\rangle$, one could also
have chosen the $|E^-\rangle$ if the {\it out}-wavefunctions are more
readily available.

The {\it second} step is the ``complex basis vector expansion".
It holds for ``very well-behaved" vectors of a subspace $\Phi_-$ of $\Phi$
(Schwartz space).  For every
$\phi^+\in\Phi_-$ (a similar expansion holds also for every
$\psi^-\in\Phi_+)$ one obtains for the case of a finite number of resonance
poles of the analytically continued $S$-matrix at the positions
$z_{R_i}=E_{R_i}-i{\Gamma_i\over2}$, $i=1,2,\ldots N$, the following basis
system
expansion:
\begin{eqnarray}
\phi^+&=&\int^{-\infty_{II}}_0d\omega|\omega^+\rangle\langle^+
\omega|\phi^+\rangle+\sum^N_{i=1}|z^-_{R_i}\rangle2\pi
\Gamma_i\langle^+z_{R_i}|\phi^+\rangle\nonumber\\
&+&\sum_n|E_n)(E_n|\phi)\quad{\rm for}\quad\phi^+\in\Phi_-
\end{eqnarray}
where $|z^-_{R_i}\rangle\sqrt{2\pi\Gamma_i}=\psi^{G_i}\in\Phi^\times_+$ are
Gamow kets (C1) representing decaying states (C2).  Their properties are
summarized in Appendix C below (see also reference 14).  The forms (B1) and (B2)
of the generalized basis vector expansions assumes that $H$ is the only
observable to be diagnalized  (cyclic operator).  If the complete system of
commuting observables (c.s.c.o.) consists of $H,B_1, B_2, \dots, B_N\equiv H,B$,
then the projection operators $|E_n)(E_n|\to\sum|E_n,b)(E_n,b|$ where the sum
extends over all values of the degeneracy quantum numbers $b$ of the energy
$E_n$.  Similarly in (B1), (B2):
\begin{eqnarray}
&&|E^+\rangle\langle^+E|\qquad\to\qquad\sum_b|E,b^+\rangle\langle^+E,b|
\nonumber\\
&&|z_R^-\rangle\langle^+z_R|\qquad\to\qquad\sum_b|z_R,^-b\rangle\langle^+z_R,
b|\nonumber
\end{eqnarray}
The operator B could be e.g. the hypercharge operator {\it if} $[H,B]=0$; it
can be the operator CP if $[H,$ CP]=0.

We will from now on omit the last
sum in (B1) and (B2), as it represents the sum over the stationary, bound states
which have no importance for the problem of this paper (no bound states
appear).  Then we have two exact but different basis vector expansions for the
same
$\phi^+$ (if we choose
$\phi=\phi^+\in\Phi_-\subset\Phi$ in (B1)):  (B1) is the standard expansion and
has a correspondence in the Hilbert space (spectral resolution of operators with
a continuous spectrum), while (B2) is new and shows that the quasi-stationary
states
$|z_{R_i}^-\rangle$ can serve as basis vectors in very much the same manner as
the stationary states
$|E_n$)
in the standard case.  But in addition to the resonance
states the new basis vector expansion (B2) also contains an integral over the
negative real axis from e.g. $E=0$ to
$-\infty_{\rm II}$ in the second sheet of the energy surface of the $S$-matrix.
This integral, called
``background term" $\phi^+_{bg}$ (which may be as much a misnomer as the term
``complex spectral resolution" for (B2)), may be
infinitesimally small, but cannot be zero. But it may also have some small but
observable consequences.  It can be calculated using the van Winter
theorem$^{14)}$ from the values
$\langle\psi^-|E^+\rangle\langle^+E|\phi^+\rangle$ for physical energies $0\leq
E<\infty$ and depends upon the apparatus for
$\phi^+$ and $\psi^-$.
\setcounter{equation}{0}
\section{Gamow vectors}

Gamow vectors are generalizations of Dirac kets, and therefore we denote them
also by kets $|\psi^G\rangle=|z_R^-\rangle\sqrt{2\pi\Gamma}$ where
$z_R=E-i{\Gamma/2}$ is the complex energy value (for every
$|z_R^-\rangle\in\Phi^\times_{+}$ there is also a Gamow vector
$|z^{*+}_R\rangle\in\Phi^\times_-,\ z_R=E-i{\Gamma/2})$.  The Gamow vectors
have the following properties:
\begin{enumerate}
\item[1.]
They are generalized eigenvectors of Hamiltonians $H$ (which we always assume to
be (essentially) self-adjoint and bounded from below) with generalized
eigenvalues $z_R=E_R-i\Gamma/2$,
\begin{equation}
H^\times|\psi^G\rangle=z_R|\psi^G\rangle
\end{equation}
where $E_R$ and $\Gamma$\ are respectively interpreted as the resonance energy
and width.
\item[2.]
They satisfy the following exponential decay law for $t\geq0$ only:
\begin{eqnarray}
W^G(t)&=&e^{-iH^\times t}|\psi^G\rangle\langle\psi^G|e^{iHt}\nonumber\\
&=&e^{-i(E_R-i\Gamma/2)t}|\psi^G\rangle\langle\psi^G|e^{i(E_R+i\Gamma/2)t}
=e^{-\Gamma t}W^G(0).
\end{eqnarray}
\item[3.]
They have a Breit-Wigner energy distribution.
\item[4.]
They obey an exact Golden Rule of which Fermi's Golden Rule is the Born
approximation.
\item[5.]
They are associated with a pole at $z_R$ in the second sheet of the
analytically continued $S$-matrix.  They are derived as the functionals of the
pole term of the $S$-matrix.
\end{enumerate}
In the absence of a vector description of resonances in the Hilbert space
formulation, the pole of the $S$-matrix has commonly been taken as the
definition of a resonance. In the RHS formulation the Gamow vectors are derived
from the pole term of the $S$-matrix$^{11)14)}$, and therefore these vectors
$|z^-_R\rangle\in\Phi^\times_+$ describe decaying resonances as
autonomous microphysical entities, in very much the same way as the $|E_n$)
describe stable particles.   (There are also Gamow vectors
$|z^{*+}_R\rangle,\ z^*_R=E_R+i\Gamma/2$, associated with the pole at $z^*_R$,
and they have an exponentially growing semi-group evolution for
$-\infty<t\leq0$).\\[6pt]
{\bf Acknowledgement}

I gratefully acknowledge helpful discussions with M. Gadella on the
mathematical properties of the background term in the basis vector expansion
(B2) and with B.  Winstein on the fits of the interference term in $R(t)$. I am
grateful to P. Kielanowski for discussions and for his help with
this paper, and to S. Wickramasekara for proofreading of the manuscript.
Part
of this paper was written at the Fundation Peyresq Foyer d'Humanisme, Peyresq,
France whose hospitality as well as financial support by NATO is gratefully
acknowledged.\\[20pt]
{\Large\bf References}
\begin{enumerate}
\item[1.]
G. Gamow, Z. Phys. {\bf 51}, 510 (1928), R.E. Peierls, {\it Proceedings of
the 1954 Glasgow
Conference on Nuclear and Meson Physics}, E.M. Bellamy, et al., editors,
(Pergamon Press, 1955);
G. Garcia-Calderon, R. Peierls, Nucl. Phys. A {\bf 265}, 443 (1976); E.
Hernandez and A.
Mondragon, Phys. Rev. C {\bf 29}, 722 (1984).
\item[2.]
T.D. Lee, {\it Particle Physics and Introduction to Field Theory}, (Harwood
Academic, Chur,
1981), and references therein.
\item[3.]
Quote from A. Pais, {\it CP-violation:
The First 25 Years}, J. Fran Thanh Van, Editor, Editions
Frontiers (1990).
\item[4.]
The real experimental situation is more complicated since the initial
neutral $K$ meson is usually not a pure $K^0$ state; but it is an
incoherent mixture of
$K^0$ and $\bar K^0$, due to the strangeness conservation in the
production
mechanism.
A. B\"ohm, P. Darriulat, C. Grasso, V. Kaftanov, K. Kleinknecht, H.L.
Lynch, C. Rubbia, H. Ticho,
K. Tittel, Nucl. Phys. B {\bf 9}, (1969), 605.
\item[5.]
P. Huet, M.E. Peskin, Nucl. Phys. B {\bf 434}, 3 (1995); J. Ellis, J.S.
Hagelin, D.V.
Nanopoulos and M. Sredicki, Nucl. Phys. B {\bf 241}, 381 (1984); J. Ellis,
J.L. Lopez,
N.E.
Mavromatos, and D.V. Nanopolous, Phys. Lett. B {\bf 293}, 142 (1992),
CERN-TH 95-99 (1995).
\item[6.]
K. Kleinknecht, {\it CP Violation}, p. 41, C. Jarskog (editor), World
Scientific (1989).
\item[7.]
J. Christenson, J. Cronin, V. Fitch, R. Turley, Phys. Rev. Lett. {\bf 13},
138 (1964).
\item[8.]
NA31, G.D. Barr, {\it et al.}, Phys. Lett. B {\bf 317}, 233 (1993).
\item[9.]
E731; L.K. Gibbons, {\it et al.}, Phys. Rev. Lett. {\bf 70}, 1199 and 1203
(1993).
\item[10.]
L.A. Khalfin, JETPh Lett. {\bf 25}, 349 (1972); L.A. Khalfin, {\it CP
violation Problem Beyond
the Lee-Oehme-Yang Theory}, DAPHNE Physics Handbook, 1994.
\item[11.]
A. Bohm, J. Math. Phys. {\bf 22}, 2813 (1981);
A. Bohm, Lett. Math. Phys. {\bf 3}, 455 (1978);
M. Gadella, J. Math. Phys. {\bf 24}, 2142 (1983); J. Math. Phys. {\bf 25},
2481 (1984); M.
Gadella, J. Math. Phys. {\bf 24}, 1462 (1983);
I. Antoniou, in
{\it Proceedings of the II.  International Wigner Symposium, Goslar}, 1991,
H.D. Doebner, {\it et al.}, Editors, (World Scientific Publishers,
Singapore, 1992); I. Antoniou,
I. Prigogine, Physica A {\bf 192}, 443 (1993);
I. Antoniou, S. Tasaki, Int. J. Quant. Chem.
{\bf 46}, 425 (1993).
\item[12.]
A. Bohm,
{\it Proceedings of the Symposium on the Foundations of Modern Physics,
Cologne, June 1,
1993}, P. Busch and P. Mittelstaedt, Editors, (World Scientific, 1993), p.
77; A. Bohm, I.
Antoniou, P. Kielanowski, Phys. Lett. A {\bf 189}:, 442 (1994); A. Bohm, I.
Antoniou, P.
Kielanowski, J. Math. Phys. {\bf 36}, 2593 (1995); A. Bohm, Phys. Rev. A
{\bf 51}, 1758 (1995).
\item[13.]
E. Roberts, J. Math. Phys. {\bf 7}, 1097 (1966);
A. Bohm, {\it Boulder Lectures in Theoretical Physics
1966}, Volume {\bf 9A}, (Gordon and Breach, New York, 1967); J.P. Antoine,
J. Math. Phys. {\bf 10}, 53 (1969); {\bf 10}, 2276 (1969); O. Melsheimer,
J. Math.
Phys. {\bf 15}, 902 (1974).
\item[14.]
For mathematical details see A. Bohm and M. Gadella, {\it Dirac Kets, Gamow
Vectors, and
Gel'fand Triplets}, Lecture Notes in Physics, Volume {\bf 348},
(Springer-Verlag, Berlin, 1989); A. Bohm, {\it Quantum Mechanics}, 3rd
Edition, Springer, NY (1994).
\item[15.]
A. Bohm, S. Maxson, M. Loewe, M. Gadella, {\it Quantum Mechanical
Irreversibility}, Physica A (1996) to appear.
(Springer-Verlag, Berlin, 1983),
\item[16.]
G. Ludwig,  {\it Foundations of Quantum Mechanics},
Volume I, (Springer-Verlag, Berlin, 1983),
Volume II, (1985).
\item[17.]
The notation $H^\times$ is the notation of
(A7).  Since $H$ is self-adjoined we usually drop the $^\times$,
i.e. we use the
notation of (A8).  The important feature of (3.2) is that
$t\geq0$ i.e. that we want to consider
$\phi^+(t)$ for $t\geq0$ and then
$\phi^+\in\Phi_-$ is
to be understood as a functional over
$\Phi_+$ i.e.
$\phi^+(t)\in\Phi^\times_+$; note
$\Phi^\times_+\supset\Phi_-\ni\phi^+$.
To emphasize this, we retain the mathematically precise notation
$e^{-iH^\times t}\equiv(e^{iHt})^\times$ in (3.2). As a consequence
of the fact that $\phi^+(t)\in\Phi^\times$ for $t>0$ it follows
that
one can take
only bra-kets
of $\phi^+(t)$
with
$\psi^-=$out-observables (like e.g. $\psi^-=|\pi\pi>$)
which means that $<\psi^-|\phi^+(t)>$ is well defined for $t>0$ but
{\it not}
$<\tilde\phi^+|\phi^+(t)>$ if $\tilde\phi^+$ is any element of $\Phi_-$.
This is sufficient for the physical problem since only
$|<\psi^-|\phi^+(t)>|^2$ has a meaning, namely the probability for a
transition from the state $\phi^+$ into the decay products $\psi^-$.
\item[18.]
Deviations from the exponential law for large values of $t$ follow from the
finite
lower limit $E=0$ in the integral (3.8) (in contrast, the integral
representation of the Gamow
vectors extends over $-\infty_{II}<E<+\infty$ where the negative $E$ are on
the second energy
sheet of the $S$-matrix).   Deviations for small values of $t$ follow from
the condition that
$\phi^+$ be in the domain of $H$ (second equation (3.10);
L. Fonda, G.C. Ghirardi, A. Rimini, Repts. on Prog. in Phys. {\bf 41}, 587
(1978)
and references thereof.
\item[19.]
Though (3.11) looks suggestive because of (3.15) it is not
obtained without
additional explanations:  Since  $\{CP,Y^{\rm op}\}=0$ the hypercharge
operator $Y^{\rm
op}$ changes the eigenvalues $cp=\pm1$ of the operator CP; thus:
\[
Y^{\rm op}|E,cp^+\rangle=\eta_{cp}(E)|E^{\prime}(-cp)^+\rangle
\]
where the phase factor $\eta_{cp}(E)$ can be chosen to unity, fixing the
relative phase of
the wave function $b_{+1}(E)$ and $b_{-1}(E)$ of (3.11) in the standard way
of (3.15). We also
choose
the phases such that:
CP$\phi^+_y=\phi^+_{-y}$  for the vector in (3.12).
However, since $[H,Y^{op}]\not=0$
(due to
$H_w$ in
$H=H_0+H_w$),  $Y^{op}$ also shifts the energy (i.e. the invariant mass of
the decay products
of
$K_{1,2}^0$) slightly.  Thus more accurately one would have
to take in place of (3.11):
$b^y_+(E)$=sign%
$(y)\ b^y_-(E^{\prime})$ where $E^{\prime}\approx E$. Since we do not want to
use the Dirac
basis vector expansion$(3.8)$ we need not pursue this point any further.
\item[20.]
The normalization of the $|z_{R_i}^{\phantom{000}-}>$
has its origin in the $\delta$-function ``normalization" of the Dirac
ket, $<^{-}E|E^{\prime-}>=\delta(E-E^{\prime})$,
and in the Titchmarsh theorem.  The ``normalization" of the
$|K_i^{\enspace-}>$ has been chosen such
that in the limit $\gamma_i/m_i\to0$ it agrees with the normalization
(unity), of the usual
$|K_{1,2}^0>$.  Since the normalization is not relevant for our problems,
we choose here the more
familiar normalization which leads to the factor $\sqrt{2\pi\gamma_i}$ in
(3.18).
\item[21.]
The Gamow vectors can be defined as vectors $\in\Phi^\times$ that are
associated with the
resonance pole of the $S$-matrix,$^{11,14)}$ in the same way as the bound
state vectors can be
defined as the vectors associated with the bound state poles of the
$S$-matrix.  Whereas in the
standard Hilbert space quantum mechanics, the bound states can also be defined
independently of
the $S$-matrix, resonance states cannot.  But in the RHS formulation both
can be defined independently
of the $S$-matrix, as (generalized) eigenvectors of $H$. Using the
$S$-matrix definition of
resonances the associated Gamow vectors can then be shown to fulfill (1.3)
and (1.4) for
$t\geq0$ (or precisely (3.7) as generalized eigenvalue equations (A6).)
The same Gamow vectors appear also in the complex ``spectral"
decomposition (B2) which is an exact representation for every very well
behaved vector
$\phi^+\in\Phi_-$ i.e. for every vector which in our interpretation
(2.1) (based on the preparation $\to$ registration arrow of
time,$^{12,22)}$) represents an
in-state prepared by an experimental apparatus.
\item[22.]
The existence of this preferred direction of time can be formulated as:  A
state must be prepared
first before an observable can be measured in it.
The Liouville equation of ref. 5 leads like our semigroup to an ``arrow of
time" however the
irreversibility of ref. 5 is due to extrinsic influences whereas in our
case  the time evolution
is generated by the Hamiltonian of the system.
\item[23.]
The Hardy class functions in the half-plane $<^\pm\omega|\phi^\pm>$ are
already determined by their
values on the positive real axis $i.e.$ by their physical values of energy
$E$, but $S(E)$ depends
upon the dynamics of the problem and $<^+E|\phi^+>$ on the choice of the
prepared state vector
(preparation apparatus).
\item[24.]
In the realistic experiment (without regeneration) one does not have pure
states but an incoherent
mixture.
But, as a consequence of strangeness conservation in the
strong production mechanism,
these are incoherent mixtures of $\phi^+_{y=+}$ $(|K^{0}\rangle)$ and
$\phi^+_{y=-}(|\bar K^{0}\rangle)$
i.e. the state is
$w_1^{(p)}|\phi_{y=1}^{+(p)}\rangle\langle\phi_{y=1}^{+(p)}|+w_{-1}(p)|\phi_
{y=-1}^{+(p)}
\rangle\langle\phi_{y=-1}^{+(p)}|$ integrated over all values of the
$K$-momenta $p$.
\item[25.]
Due to $[H,Y^{op}]\not=0$ the operator $Y^{op}$ is not a symmetry group
generator but a spectrum generating group generator changing not only the
eigenvalue  cp of
CP but also the eigenvalue $z_i$ of $H$:
\begin{eqnarray}
Y^{op}|z_1,cp=1,^-\rangle&=&\eta_1|z_2,cp=-1,^-\rangle\nonumber\\
Y^{op}|z_2,cp=-1,^-\rangle&=&\eta_{-1}|z_1,cp=-1,^-\rangle\nonumber
\end{eqnarray}
where $\eta$ are phase factors (which can be chosen $+1$ by absorbing then
into the
$b_i$:  $b_1\to\eta_1b_1\quad b_{-1}\to\eta_{-1}b_{-1}$).
These relations are needed to establish (3.25).  Cf. also footnote 19.
\item[26.]
L. Maiani (editor) The DA$\Phi$NE  Physics Handbook,
Vol. 1 (1992); R. LeGac, Study of CP, T and CPT in the neutral K system.
Workshop on K Physics, Orsay, May 1996; ref. 5 above.
\item[27.]
L.K. Gibbons, University of Chicago Dissertation, Chicago, August 1993
(in particular sect. 9.2 and 9.3) C. Geweniger, {\it et al.}
Phys. Lett. 48B, 487 (1974); J.H. Christenson {\it et al.}
Phys. Rev. Lett {\bf 43}, 1212 (1974) and ref. 8 and 9 above.
\item[28.]
To be more scrupulous we should calculate the time dependent transition rate
${\cal P}(t)$ using the exact Golden Rule of RHS quantum mechanics,$^{14)}$
which is possible and will be done in a future publication.
But in order not to complicate the presentation
further by novel arguments, which are not relevant for
the problem under consideration, we use this semi-heuristic $R(t)$
and the observable parameters $\eta_{+-}$
(and $\eta_{00}$) defined
in terms of the effective theory for the phenomenological
analysis of our exact (untruncated) theory (3.27).
\item[29.]
The observable parameters $\eta_{+-}$ (and $\eta_{00}$) are by
(1.10) defined in terms of
the effective theory.  But they are experimentally determined as the ratio
of the
$K^0\to\pi^+\pi^-$ rate at long times $t\approx20{1\over\gamma_S}$ to the
rate at short
times ($t={1\over\gamma_S}$ extrapolated back to $t=0$), when the
$K^{0}\rightarrow\pi^+\pi^-$
are mainly due to $K_S\rightarrow\pi^+\pi^-$.  Thus
the value (1.12) for
$\eta_{+-}$ is also in the present case related by (1.11)
with sufficient accuracy to
$R(t=20\tau_S)$.
\end{enumerate}
\newpage\noindent%
{\large\bf Figure caption}\\[4pt]
{\bf Figure 1}: The definition of the spaces $\Phi_-$ (in-states) and
$\Phi_+$ (out-observables).
The
preparation and registration procedure for quantum
systems$^{16)}$ is applied to the inelastic scattering experiment for production
and decay of the neutral $K$-system.   Fig. 1a shows the preparation of the
$\pi^-$ state.  Fig. 1b depicts the preparation apparatus of the
$K^0$ system which consists of the preparation of $\pi^-$
and the interaction with the prepared baryon system B.  Fig. 1c
shows
the registration apparatus which defines
the out-``state" $|\psi^-\rangle\langle\psi^-|$
(observable);
it principally consists of the $\pi\pi$
detector.  Every arrangement for an experiment with single microsystems consists
of a preparation apparatus and a registration apparatus.  Fig. 1d shows how the
preparation part of Fig. 1b, is combined with the
registration part of Fig. 1c into the experiment that
measures the probability for the transition $K^0\to\pi\pi$.  Since the
in-states $\phi^+$ and the out-observables $\psi^-$ are subject to different
conditions, they are described in the RHS quantum theory by distinct spaces
$\Phi_-$ and $\Phi_+$, respectively.
\begin{figure}
\epsfxsize=\textwidth
\epsffile{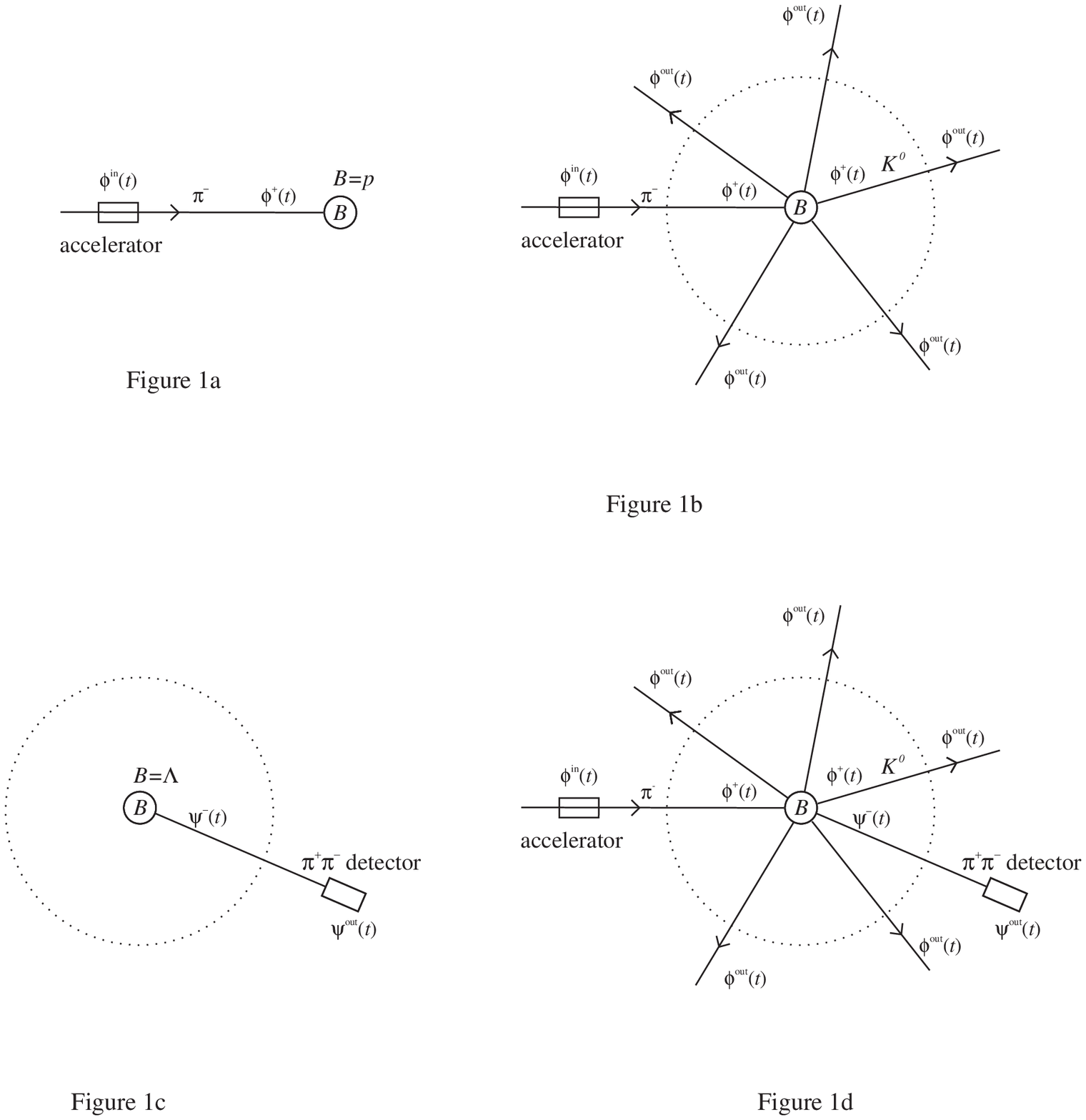}
\end{figure}
\end{document}